\documentclass[pre,aps,preprint,floatfix]{revtex4}
\usepackage{natbib}
\usepackage[cp850]{inputenc}
\usepackage{amsmath,amssymb,color}
\usepackage{graphicx}
\usepackage{latexsym}

\newcommand{\be}{\begin{equation}}
\newcommand{\ee}{\end{equation}}

\newcommand{\fr}{\displaystyle\frac}

\begin{document}

\title{Liquid flow-focused by a gas: jetting, dripping and recirculation}
\author{Miguel A. Herrada, Alfonso M. Ga\~{n}\'an-Calvo, Antonio Ojeda-Monge,
Benjamin Bluth, Pascual Riesco-Chueca}
\affiliation{E.S.I,
Universidad de Sevilla.\\
Camino de los Descubrimientos s/n 41092 Spain.}
\date{\today}
\begin{abstract}
The liquid cone-jet mode can be produced upon stimulation by a
co-flowing gas sheath. Most applications deal with the jet
breakup, leading to either of two droplet generation regimes:
jetting and dripping. The cone-jet flow pattern is explored by
direct axisymmetric VOF numerical simulation; its evolution is
studied as the liquid flow-rate is increased around the
jetting-dripping transition. As observed in other focused flows
such as electrospraying cones upon steady thread emission, the
flow displays a strong recirculating pattern within the conical
meniscus; it is shown to play a role on the stability of the
system, being a precursor to the onset of dripping. Close to the
minimum liquid flow rate for steady jetting, the recirculation
cell penetrates into the feed tube. Both the jet diameter and the
size of the cell are accurately estimated by a simple theoretical
model. In addition, the transition from jetting to dripping is
numerically analyzed in detail in some illustrative cases, and
compared, to good agreement, with a set of experiments.
\end{abstract}
\maketitle

\section{Introduction}

The controllable production of small flowing geometries is a
crucial challenge for chemical engineering \cite{Basaran02} and
bio-industry \cite{bio,grif}. Drops, bubbles, jets and
recirculation cells in the micro-scale provide a useful platform
for diverse technical applications. Here we concentrate on
cone-jet flow patterns, looking at the streamline geometry in the
jetting to dripping transition. Recirculating flow is shown to
take place just before the transition. This may have been
disregarded in some experimental setups, whose prime concern was
the study of the cone and jet geometry, or the analysis of drop
generation. Its study requires either specific flow visualization
techniques or numerical simulation methods. The conditions for
recirculation are extremely interesting, both as an indicator
phenomenon associated to the jetting-dripping threshold, and as an
attractive technological feature.

Small droplet generation by means of co-flowing immiscible fluid
streams has become widespread. The intrinsic smallness of the
output droplets generally leads to small Reynolds number flows.
Thus, a number of classic studies back to Taylor \cite{Taylor34}
including the recently blooming field of co-flowing microfluidics
take the low Reynolds number assumption for granted. For example,
a simple scheme (a straight tube surrounds a coaxial, more slender
tube, two immiscible fluids being fed through each tube) has been
thoroughly explored by Suryo and Basaran \cite{SB2006} using a
computer simulation: a locally extensional flow spontaneously
develops at the tip of the elongated drop drawn by the co-flowing
liquid, causing the seemingly continuous ejection of extremely
small droplets by tip streaming under certain parametrical
combinations. This is obtained without the burden of complex
geometry. It is hoped, therefore, that low Reynolds number
co-flowing small-droplet generation may become a hydrodynamic
standard in a near future. Different setups have been proposed
where the flow is driven by an external straining flow. Among
them, the elegant analytical solution by Zhang \cite{Wendy04}
points to parametrical combinations where extremely thin fluid
jets, even down to the molecular scale, could be continuously
reached. Those jets, if confirmed, would yield unimaginably small
droplets upon breakup.

An interesting research field is concerned with the behavior of
electrified cones and drops of leaky dielectric fields. The
problem of freely-suspended liquid droplets deforming due to an
applied electrostatic field was examined by Haywood, Renksizbulut
and Raithby \cite{hayw}. Collins et al. \cite{Bas07} reported
simulations and experiments supplying a comprehensive picture of
the mechanisms of cone formation, jet emission and break-up that
occur during EHD tip streaming from a liquid film of finite
conductivity. Lac and Homsy used a boundary integral method to
describe the axisymmetric deformation and stability of a viscous
drop in a steady electric field \cite{Lac}. In the present paper,
however, only non-electrified fluids will be considered.

In spite of their generality and tractability, low Reynolds number
flows are constrained by the requirement that the overall flow
velocity does not grow above a certain threshold to ensure that
inertial forces remain negligible. This constraint limits the
overall productivity of low-Re systems. Co-flowing with inertia
was successfully explored, aiming at a reduction of the issued
bubble diameter, by Oguz and Prosperetti\cite{OP93}. Subsequently,
new perspectives were open by the emergence of moderate-high
Reynolds number flow-focusing \cite{Gan98} as a high-productivity
alternative to low Reynolds number co-flowing systems. Compared to
other co-flowing techniques, flow focusing (FF) stands today as a
mature microfluidic standard yielding steady capillary jets or
droplets whose size is well below the scale of the flow
boundaries. As originally conceived, FF aimed at the generation
\cite{Gan98} of continuous steady micro-jets upon focusing by a
co-flowing gas stream at moderate-high Reynolds number.
Furthermore, FF was shown \cite{Gan01} to produce perfectly
monodisperse microbubble streams when the co-flowing current is a
liquid. A slight variation of the concept was subsequently
introduced by Takeuchi et al. \cite{Take05} to produce
microbubbles. When the axisymmetric geometry originally proposed
was reduced to a planar topology \cite{Anna03}, particularly
suitable for microfluidics, the scientific literature production
on flow focusing underwent an enormous
boost\cite{Garstecki05,Seo05,Jensen06}. In addition, axisymmetric
multiple-phase FF leading to compound co-axial
microjets\cite{Gan98}
has been developed by other authors to produce microcapsules in a
microfluidic setup at relatively low \cite{Utada05} and moderate
Reynolds number\cite{Berk04}.

The technological applications of FF were evident from its very
inception. A crucial advance resulted from the combination of
massive production (high production rates of microscopic fluidic
entities) and accurate tailoring. Depending on the geometry and
arrangement of the involved fluids (a decision determining which
interfaces are to be created), the nature of the fluids involved,
gas or liquid, and the system geometry, an output including nearly
monodisperse micro-droplets, bubbles or complex capsules can be
obtained at an unprecedentedly controllable rate. Surface tension
becomes a paramount ally in the conformation of discrete
(generally spherical) fluidic units. Capillary jets have long ago
been observed to give rise to continuous drop streams at fast
emission rate upon Rayleigh axisymmetric breakup. Here, although
surface tension is negligible compared to other driving forces in
the global scale, it becomes the main driving agent for jet
instability and breakup. Obtaining a jet is therefore the
precondition for the creation of a fluid domain with higher
velocities and smaller dimensions at no cost in terms of control;
and surface tension is free to perform its conformation task in
this new scale. Thus, as first proposed in FF, the steady
capillary thin jet conformed by pressure forcing by an immiscible
co-flowing fluid provides favorable local conditions, a suitable
environment for the generation of bubbles, capsules or droplets.

A FF capillary jet is driven by three main agents: fluid inertia,
viscosity and surface tension. Owing to the simplicity of the
slender jet geometry, which asymptotically renders all forces
strictly additive in one dimension, FF can be scaled with the help
of two dimensionless parameters: (i) the inertia to surface
tension forces ratio (Weber number) and (ii) the viscous to
surface tension ratio (Capillary number). Other classic numbers
such as Reynolds are combinations of the former. Nevertheless, as
early noticed \cite{Gan98}, an intrinsic feature of FF, namely the
presence of a focusing fluid, gives rise to supplementary
influences issuing from the correlation between the properties of
the focusing and focused fluids. In particular, when a liquid is
being focused by a gas, the gas sheath flows much faster than the
liquid jet at the exit orifice (Fig. 1). Thus, in addition to the
extensional viscous forces at the neck of the meniscus,
transversal viscous diffusion of momentum causes a non-trivial
axial velocity profile. Some simplifying assumptions have been
adopted, yielding accurate first order solutions
\cite{Gan98,small05,advmat06}. However, they are not applicable to
predict critical phenomena like the onset of steady jetting, or
the jetting-dripping transition, as a function of the working
parameters. These problems have been made analytically tractable
at the expense of a drastic geometry simplification, i.e. assuming
infinite jet slenderness \cite{AP06,Gan06,Gan07}. Such simplified
models are predictive in a variety of situations, but FF systems
exhibit an intrinsically three-dimensional meniscus from which the
jet or the small droplets issue. Simultaneous modelling of the
meniscus and the jet goes beyond the scope of present theoretical
frameworks. Thus, numerical simulation or experiment are the only
avenue to discern the physics of the fluid emission and its
parametric conditions. Some further insight can be gained by
general scaling laws. This is the approach chosen for the research
presented here.

Many authors have applied numerical simulation \cite{Erickson05}
to this class of problems, where it has supplied welcome
information on the droplet dynamics of complex flows
\cite{Cristinni2004}. A significant number of studies have been
proposed on microfluidic FF devices; occasional comparison with
experimental data is provided to validate the numerical model. A
liquid-liquid configuration for the production of microemulsions
has been simulated \cite{Dupin06} to good agreement with
experiments \cite{Anna03}. Other authors have considered the
microbubbling setup \cite{Jensen06,Weber07}, where good
experimental fit is also obtained\cite{Garstecki05}.

In this work, we make use of numerical simulation, with some
experimental support, to study the generation of a liquid jet
focused by gas in an axisymmetric FF device, at moderate-high
Reynolds numbers. The jet diameters obtained in the simulation are
in good agreement with our experiments and scaling laws
\cite{Gan98}, a fact that fully validates our hypotheses. Among
other findings, we determine the flow rate at the jetting-dripping
transition for two combinations of Reynolds and Weber numbers. We
observe that either the jet or the cusp-like meniscus are
responsible for the global instability of the system, which drives
it into well defined dynamical cycles (global dripping). A
detailed description of the flow pattern sheds light on the
physics of the jetting-dripping transition and the peculiar
appearance of these two regimes in co-flow problems, as opposed to
faucet jetting and dripping.

One of the key findings of the simulation is the occurrence, under
favorable driving conditions, of a recirculation cell in the
meniscus. This is in perfect analogy to recirculating meridian
fluid flows observed inside Taylor cones when electrospraying
liquids with sufficiently large values of both the viscosity and
the electrical conductivity \cite{BG98,selfr}. Additionally, in
experiments aiming at the production of tip streaming patterns in
liquid-liquid two-dimensional FF (surfactant treated interface),
the streamline image of fluorescent particles seeding the flow of
the internal, aqueous liquid during thread formation, was shown to
consist of symmetric recirculation vortices \cite{AnnaMayer06}. In
both types of motion, either driven by the electrical stresses
acting at the cone surface or by the external focusing flow, the
liquid flows towards the meniscus (cone) tip, along the
generatrix, and away from it along the axis. The problem under
consideration here is comparable to these other instances of
recirculating cell, because the driving action of the gas sheath,
which causes a strong tangential forcing at the interface, plays a
similar role to either tangential electric
stress\cite{BG98,selfr}, surfactant-aided liquid-liquid
interaction\cite{AnnaMayer06}, or purely surfactant-driven
tip-streaming.\cite{Homsy1,Homsy2} In this work, scaling arguments
are developed to describe the size and occurrence of purely FF
recirculation cells.

\section{Governing equations and boundary conditions}

\begin{figure}
\begin{center}
\resizebox{0.9\textwidth}{!}{\includegraphics{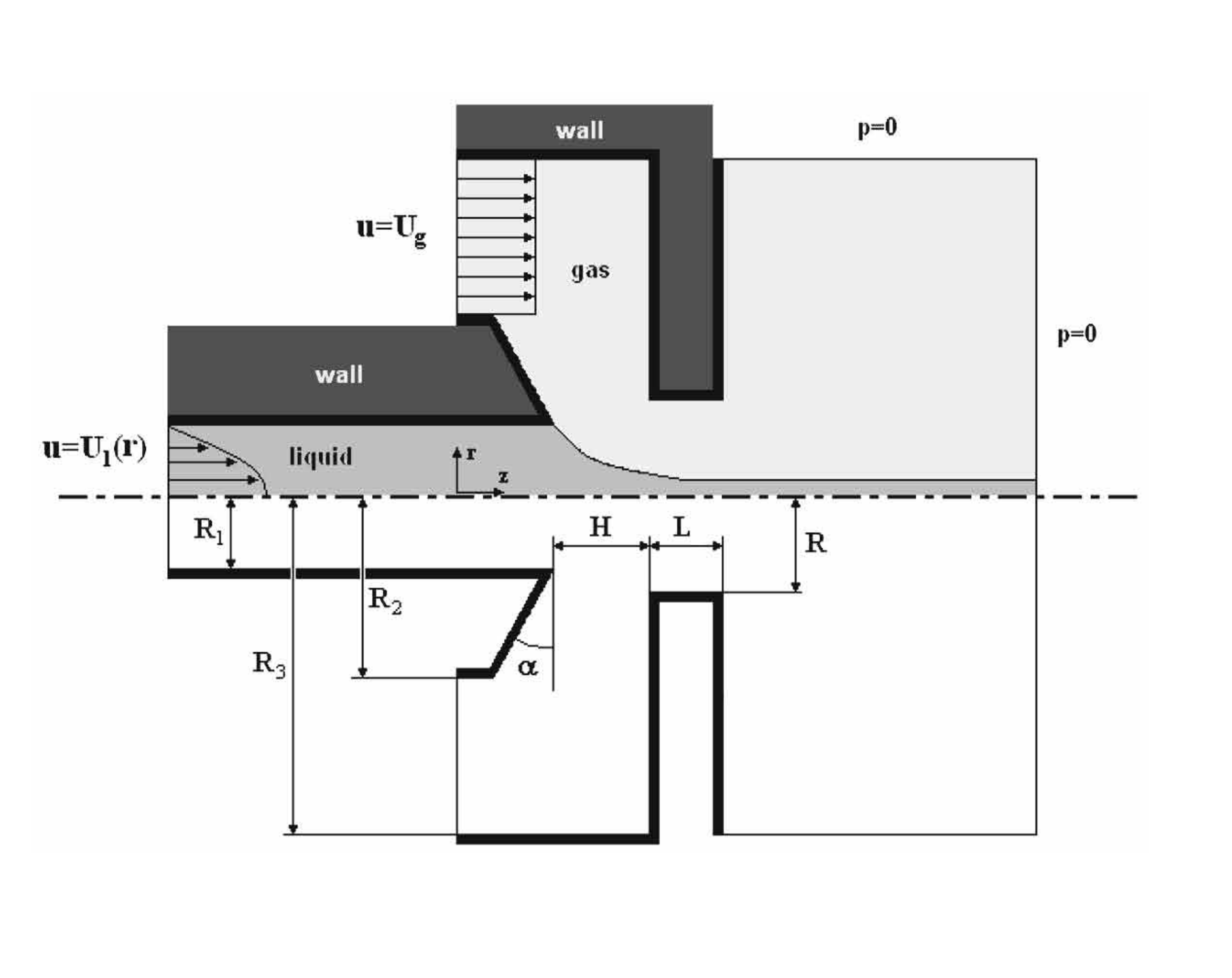}}
\end{center}
\caption{Simulated boundary geometry and fluid flow domains.}\label{f2-1}
\end{figure}

The axisymmetric flow-focusing device and the computational domain used are sketched in Fig.
\ref{f2-1}. A constant liquid flow rate $Q_l$, flowing through a capillary tube (outer diameter
$D_2=2R_2$, inner diameter $D_1=2R_1$), is forced through a coaxial round orifice of diameter
$D=2R$ (nozzle) located at an downstream distance $H$ from the tube outlet. The liquid stream is
drawn by a constant flow rate $Q_g$ of focusing gas stream discharging through the nozzle into a
infinitely large chamber. The gas flow is assumed incompressible, in asymptotic consistency with
the low pressure drop at the exit orifice, a condition prevailing when maximum droplet size
monodispersity is required. Therefore, the incompressible, axisymmetric and unsteady Navier-Stokes
equations in cylindrical $(z, r, \phi)$ coordinates are used to describe the time evolution of both
fluids.

Fig. 1 also shows the boundary conditions: a) at the liquid inlet,
$z=-z_l$, a Hagen-Poiseuille profile, $U_{l}(r)=V_1[1-(r/R_1)^2]$
, is specified; b) at the gas inlet, $z=0, R_{2}< r <R_{3}$, a
uniform axial flow, $U_{g}(r)=V_2$, is assumed. This assumption is
based on the following: the gas Reynolds number under the
conditions explored is relatively high (of the order of, or above
$100$), while most flow-focusing devices have a gas inlet length
$L_{g}$ which is not much bigger than the width $\triangle R =
R_{3} -R_{2}$, so that the relevant dimensionless number for
boundary layer development, $\rho_{g} U_{g} (\triangle R)^{2}/
(\mu_{g} L_{g})$ is sufficiently above unity; c) on all solid
walls we assume no slip and no penetration
$\mathbf{u}=\mathbf{0}$; d) at the axis $r=0$ a symmetry condition
is applied; e) the outlet discharge chamber has been modelled as a
rectangular box, $z=z_{out}$ and $r=r_{out}$ being two open
surfaces where the pressure is set to zero. This assumption is
discussed later on.

Note that the corresponding gas and liquid flow rates can be
derived from the inlet velocity field:

\be Q_l=\int_0^{R_1} 2\pi r U_l(r) dr;\quad Q_g=\int_{R_2}^{R_3}
2\pi r U_g(r) dr. \ee

Parametric studies of the dimensionless variables involved are
carried out next. The velocity field $\mathbf{u}=(u,v)$ is scaled
with the mean gas velocity at the nozzle $V=Q_g/(\pi R^2)$, while
length is scaled with the nozzle radius $R$, time $t$ with $R/V$,
and pressure $p$ with $\rho_{g} V^2$, $\rho_{g}$ being the density
of the focusing gas. A single geometrical configuration is
considered in this work, characterized  by the following aspect
ratios: $R_1/R=0.75$,$R_2/R=1.75$, $R_3/R=3.5$, $H/R=1$ and
$L/R=0.75$. We have chosen a liquid-gas combination where
 \be\fr{\rho_l}{\rho_g}=833.33, \fr{\mu_{l}}{\mu_g}=55.55 \quad\ee
$\rho$ and $\mu$ being the density and viscosity of the liquid
(subindex $l$) and the gas ($g$). This choice is representative of
the experimental jetting of air-focused water. The problem is
governed by the following dimensionless parameters: Reynolds and
Weber numbers \be Re=\frac{\rho_{g} VR}{\mu_g}, \ee

\be We=\frac{\rho_{g} V^2R}{\sigma};
 \ee
$\sigma$ being the surface tension between the two phases. $Q$ is
defined as the flow rate quotient:

 \be
Q=Q_l/Q_{g}. \ee

For a given value of $Re$ and $We$ we wish to analyze the formation of a steady liquid jet and the
dependence of the flow on the quotient $Q$. In particular, we identify the minimum value of $Q$,
$Q^*(Re, We)$, below which the liquid jet ceases to be steady and a dripping regime is observed in
the simulation. The regime is considered to be steady (and the jet convectively unstable) if the
liquid meniscus remains steady for a sufficiently large period of time.

We should point out that in order to focus a jet of liquid by gas, moderate-high Reynolds
numbers are needed. We consider in detail two different conditions for the focusing gas:

 Case 1: $Re=465.83$, $We=8.137$.

 Case 2: $Re=931.666$, $We=32.55$.

Each case will be explored under different flow rate quotients.

\section{Numerical procedure}

In order to predict the interface geometry during the
time-solution, several techniques have been used, falling into one
of three categories. These are: (i) interface tracking methods,
including a moving mesh, \cite{interf} (ii) front tracking and
particle tracking schemes, \cite{front} and (iii) interface
capturing methods, including volume of fluid (VoF)\cite{vof1,vof2}
and level set techniques.\cite{levelset} We chose a VoF method
consisting of two parts: an interface reconstruction algorithm to
approximate the interface from the set of volume fractions and a
VoF transport algorithm to determine the volume fraction at the
new time level from the velocity field and the reconstructed
front. The basic method is robust and flexible, and is based on
widely used VoF schemes
\cite{Hirt81,Heinrich91,Tomiyama93,Lafaurie94}.

\begin{figure}
\begin{center}
\resizebox{0.9\textwidth}{!}{\includegraphics{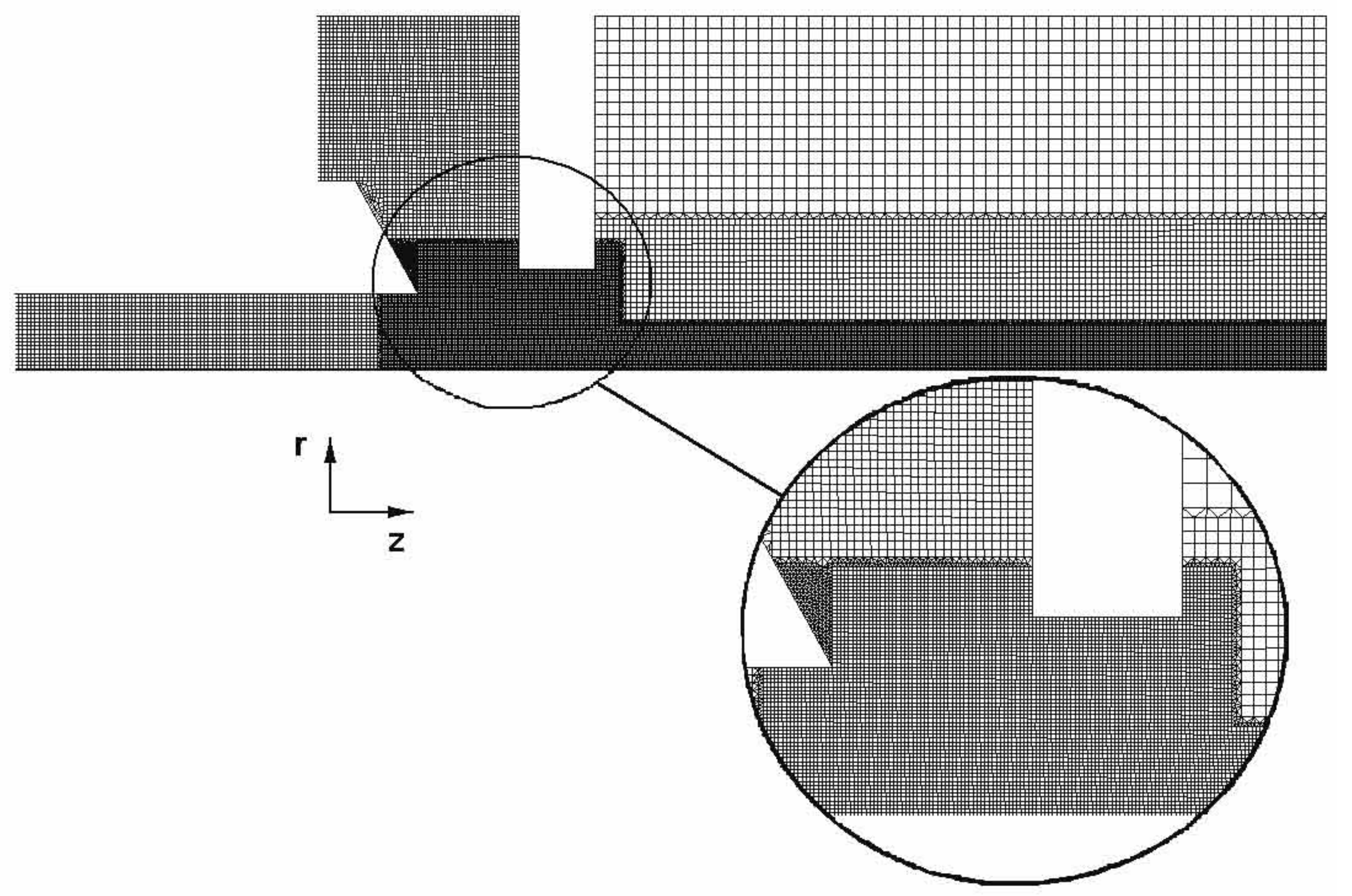}}
\end{center}
\caption{Grid of the domain under study. A denser mesh is provided
in areas where the interface is expected to lie. To avoid
numerical diffusion of the interface, the interface region is
defined with a higher density of nodes.} \label{f2-2}
\end{figure}

For convenience and with the aim of making our results readily
reproducible for others, we have used the well tested commercial
solver FLUENT v 6.3 (laminar unsteady) to resolve the discretized
mass continuity, momentum conservation, and the liquid volumen
fraction equations in the mesh depicted in Fig. 2, generated by
commercial code GAMBIT in FLUENT v 6.2. Observe that the smallest
quadrilateral elements lie between the needle edge and the nozzle,
where the liquid meniscus is located, and in the near-axis region,
where we expect the development of the liquid jet. The basic mesh
should be sufficiently refined to capture, in the absence of the
liquid, the strong velocity gradients experienced by the gas flow
at the orifice region. In the grid shown in Fig. 2 the minimum
cell radial and axial lengths are $(\Delta
z)_{min}=(\Delta_r)_{min}=0.02$. Several numerical tests with
smaller size mesh cell have shown that this level of accuracy is
comfortably sufficient to describe the gas flow pattern for the
two cases considered ($Re=465.83$ and $931.666$). All results
presented here were initially computed in that mesh. In all
instances where Q was very small, the results were recomputed in a
refined mesh with quadrilateral cells in the nozzle and jet
region, with $(\Delta z)_{min} = (\Delta r)_{min} = 0.01$.
Finally,  only for the more difficult cases (case 2 with $Q$
small), the results where recomputed in a finer grid with $(\Delta
z)_{min} = (\Delta r)_{min} = 0.005$.

A factor requiring consideration is the location of the outlet
boundaries. They are to be sufficiently remote from the nozzle to
avoid the numerical reflection of pressure waves, since the
pressure is artificially kept fixed at these boundaries during the
time evolution of the flow. Moreover, the artificial boundary
condition causes problems when the jet flows through the outflow
boundary, and more acutely, when a string of drops flows through
it. When a jet extends all the way to the outlet boundary of the
flow domain, the pressure within the jet and the surrounding fluid
cannot be equal owing to surface tension. However, for our
particular problem, as will be shown later, these undesired
effects are confined to a length below two diameters upstream of
the boundary. Therefore, we have chosen simulations with a
sufficiently large external chamber, $z_{out}=10$ and
$r_{out}=3.5$, to minimize artificial boundary effects in the
results obtained. On the other hand, the $z$ position where the
inlet boundary conditions for the liquid are imposed, has been
located sufficiently far away from the needle edge, at $z=z_l=-3$.
This choice has been made, as will be shown later, because a
liquid recirculation cell intrudes upon the capillary tube when
$Q$ decreases. Therefore, in order to impose a Hagen-Poiseuille
profile for the liquid velocity as a well posed inlet boundary
condition, this boundary should be set sufficiently far upstream
from the recirculating region.

Tracking the interface between the phases is accomplished by
solving a continuity equation for the volume fraction of one of
the phases using an explicit time-marching scheme. The rest of the
equations are solved implicitly. The time steps selected were
fixed and sufficiently small to ensure that the global Courant
number based on the mesh cell size, the mean velocity in the cell
and the time step was always smaller than one. Regarding the
spatial discretization of the equations, the third-order modified
MUSCL scheme\cite{MUSCL} is used to obtain the face fluxes
whenever a cell is completely immersed in a single phase. When the
cell is near the interface, the CICSAM algorithm is
used\cite{Ubbink97}. The pressure corrections are computed with
the body forces weighted scheme and the pressure-velocity coupling
in segregated solver is treated with the PISO method \cite{PISO}.
All under-relaxation factors are set to one to avoid any numerical
masking of fade-out effects in our physical problem.

\section{Numerical results. Discussion}

A fruitful interpretation of the results obtained needs to be
situated in the frame of the literature on the dripping faucet.
The book by Shaw \cite{shaw} gave rise to a rich and insightful
series of studies, among them major contributions by Fuchikami et
al. \cite{fuchi}, Ambravaneswaran and co-workers \cite{Ambra00,
Ambra04, Ambra06}, and Coullet, Mahadevan and Riera.\cite{Co05} To
discriminate between the jetting and dripping modes, it is helpful
to make use of the categories introduced by Ambravaneswaran and
co-workers: \cite{Ambra04}

\begin{enumerate}
 \item  the dimensionless limiting length $L_{d}/R_{1}$ from the capillary
edge to the extremity of the first drop at detachment.

 \item  the ratio of the distance $L_{s}$ between the centers of mass of
the drop that is about to form and the previously formed drop, and
$L_{d}$

 \item  the ratio of the volume of the drop that is about to form,
$V_{d}$, to that of the drop that is attached from the capillary,
$V_{p}$
\end{enumerate}

When undergoing the transition from dripping to jetting, the first
parameter undergoes a sudden increase, while the two other ones
experience an opposite trend. In general, the dripping mode is
characterized by bulky drops, relatively distant from each other,
whose diameter is considerably larger than the jet diameter from
which they detach.

The usual categories applicable to faucet dripping need some
adaptation before being used in a co-flow problem, where there is
considerable stretching of the cone-jet and droplet train by the
coaxial current: the drops are deformed, their radial extent is
limited, and they may undergo secondary breakup (particularly so
in the dripping regime, whose bulky drops are more vulnerable to
shear) after detachment from the filament. Therefore, under
co-flowing, the classical aspect of the jetting and dripping
regimes is modified, and the transition between them is not sharp:
such features are confirmed by experiment, as explained below.
This is the reason why the behavior of the meniscus can be used as
a further indicator of the jetting mode. Dripping leads to a
pulsating meniscus, each detached drop giving rise to recoil and
oscillating; while, in jetting, the detachment of the drops does
not cause any fluctuation of the meniscus and jet (see Fig.
\ref{f2-16new1}). In a full dripping regime, these pulses are
perfectly regular (see Fig. \ref{f2-13}); in most cases a slender
unsteady liquid ligament detaches from the meniscus and breaks up
into droplets of heterogeneous size \cite{Vill04}. However, at the
onset of dripping (a situation which will be labelled ``incipient
dripping''), completely irregular fluctuations of the meniscus are
observed \cite{Ambra04}.

\begin{figure}
\centerline{\includegraphics[width=0.9\textwidth]{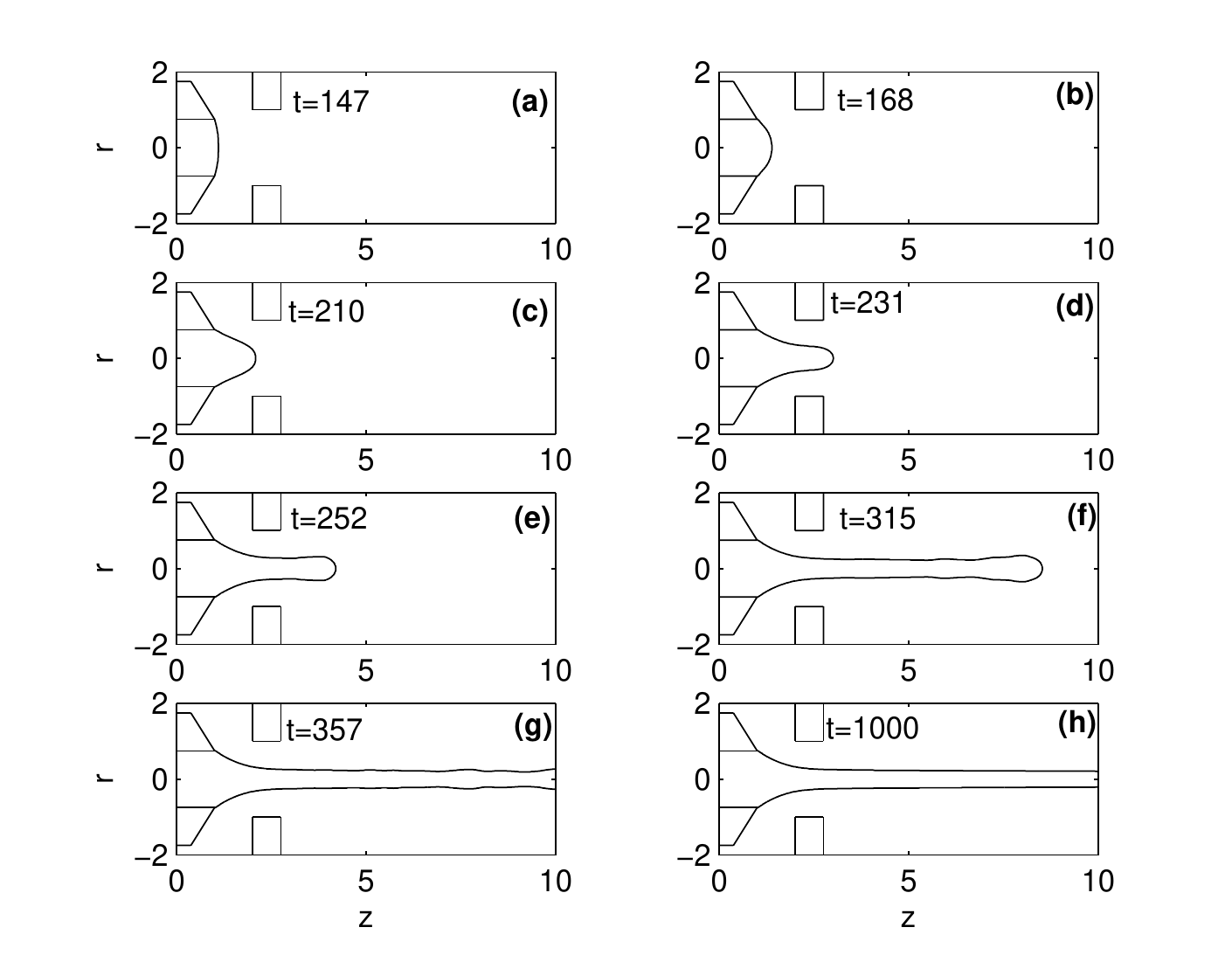}} \caption{A sequence of snapshots from
the simulated growth of an eventually steady jet (case 1, $Q=0.004 $): (a) the interface arrives to
the needle edge; (b)-(c) the  meniscus grows in the nozzle region; (d)-(f) a jet begins to issue
from the nozzle; (g)-(h) the meniscus-jet system is steady.} \label{f2-3}
\end{figure}

Accordingly we begin by studying the formation of a steady (convectively unstable) liquid jet in
the FF device. Initially, the capillary needle is filled with liquid up to $z=0$ while the rest of
the domain is filled with gas. We start the simulation from rest ($u=v=0$) in the whole domain
except at the inlet sections, where velocity profiles are prescribed. Fig.
\ref{f2-3}(a)-\ref{f2-3}(h) shows the formation of a steady liquid jet for case 1 and $Q=0.004$,
going through the stages of interface entry, meniscus growth and jet consolidation. The shape of
the liquid-gas interface is computed in the figure as the iso-level of the liquid volumen fraction
$\alpha=0.999$, obtained with the VOF method. Given that the flow might be unsteady, we consider
that the meniscus-jet has reached a steady condition whenever two conditions hold:
\begin{enumerate}
 \item  The angle between the liquid meniscus and the radial coordinate at the
capillary needle, $\theta(t)$, has reached a constant value in time, $\theta(t)=\theta_o$.

\item Both the jet diameter at the nozzle inlet, $d_{in}(t)$, and
at the nozzle exit, $d_{out}(t)$, should reach a steady regime or a stable oscillating regime
around a mean value; these quantities are of course to stay above zero. This amounts to excluding
jet breakup in the nozzle region, a feature associated with a non-slender jet and possible dripping
behavior (unsteady meniscus-jet).
\end{enumerate}

Here, $d_{in}(t)$ and $d_{out}(t)$ are computed at each time step,
by integrating radially the liquid volume fraction, $\alpha$, at
the nozzle inlet, $z=2$, and at the exit, $z=2.75$:

\be d_{in}(t)=2\sqrt{2\int_0^1 \alpha(t,z=2,r) r dr}; \quad d_{out}(t)=2\sqrt{2\int_0^1
\alpha(t,z=2.75,r) r dr}. \ee

For sufficiently large $Q$, as illustrated in Fig. \ref{f2-3}, both $d_{in}$ and $d_{out}$ evolve
towards a steady value.

\begin{figure}
\centerline{\includegraphics[width=0.6\textwidth]{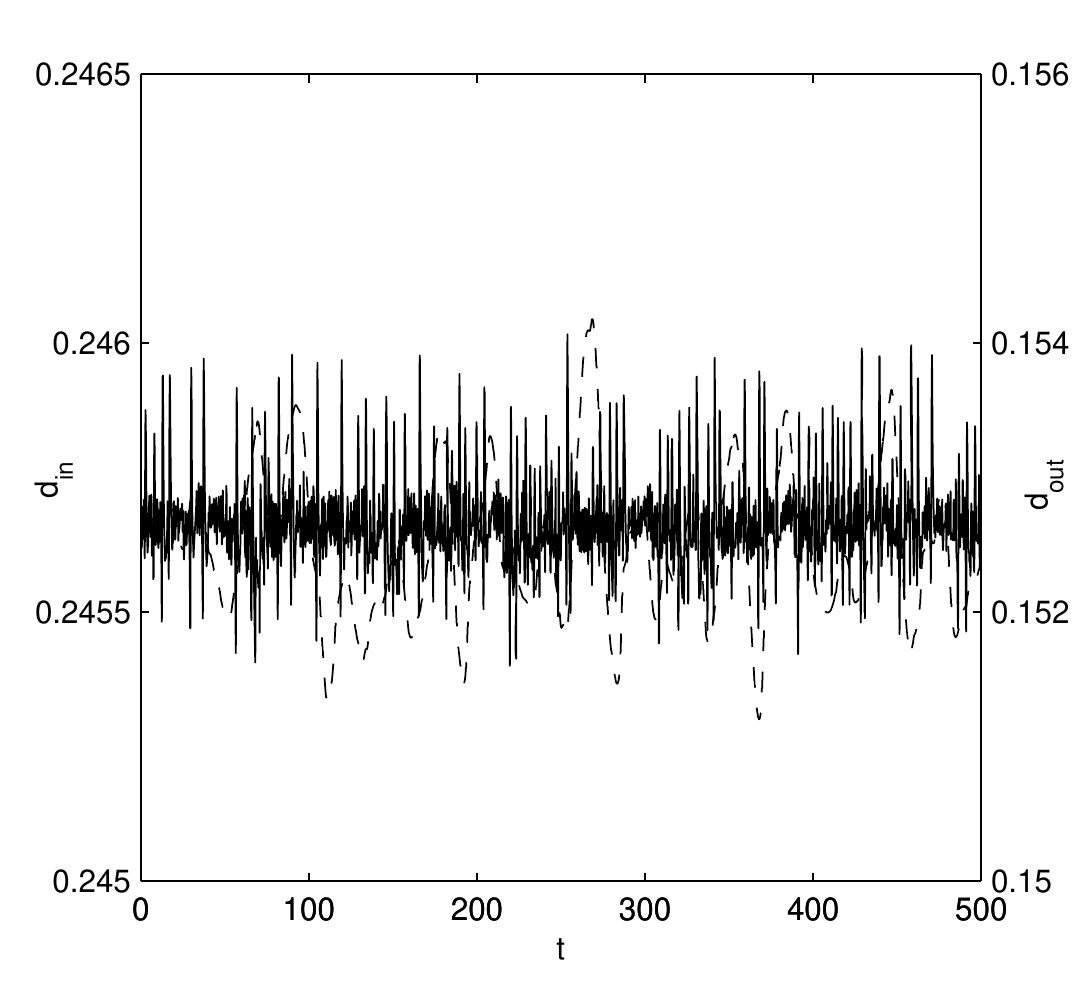}} \caption{Time dependence of the jet
diameter, $d_{in}$ (dashed line, left ordinates) and $d_{out}$ (continuous line, right ordinates)
for case 1 and $Q=0.0006 $. Note (i) the smallness of the oscillation amplitudes and (ii) the
higher frequency in the variation of $d_{out}$ as compared to $d_{in}$.} \label{f2-4}
\end{figure}

However, oscillations of these two quantities are observed when $Q$ is reduced. For example, Fig.
\ref{f2-4} shows the time oscillation of $d_{in}$ and $d_{out}$ for case 1 and $Q=0.0006$ after
allowing a steady jet to develop. It can be observed that the jet diameter at the nozzle exit is
smaller than at the inlet; mass conservation arguments imply the inlet velocity to be smaller than
the outlet velocity (in inverse proportion to the diameter squared). This explains why the
oscillation frequencies of the jet diameter are shorter at the outlet. Although the oscillation of
the jet in the nozzle region may play an important role in the dynamics of the droplets generated
upon jet breakup, our main concern here is to characterize the jet diameter, the angle of the
meniscus at the attachment, $\theta$, and the flow structure inside the meniscus as a function of
$Q$. Since the flow is unsteady, we will use a mean value of the jet diameter at the nozzle inlet
and outlet as defined by: \be \bar{d}_{in}=\frac{1}{T}\int_{t_i}^{T+t_i} d_{in}(t) dt,\quad
\bar{d}_{out}=\frac{1}{T}\int_{t_i}^{T+t_i} d_{out}(t), dt \ee where $t_i$ is a time position once
a steady jet has developed and $T$ is a time period long enough to ensure a significative mean
value. For example, selecting $t_i=0$ and $T=500$  leads to $\bar{d}_{in}= 0.2456 $ and
$\bar{d}_{out}=0.1527 $ (conditions as in Fig. \ref{f2-4}).

\begin{figure}
\centerline{\includegraphics[width=0.9\textwidth]{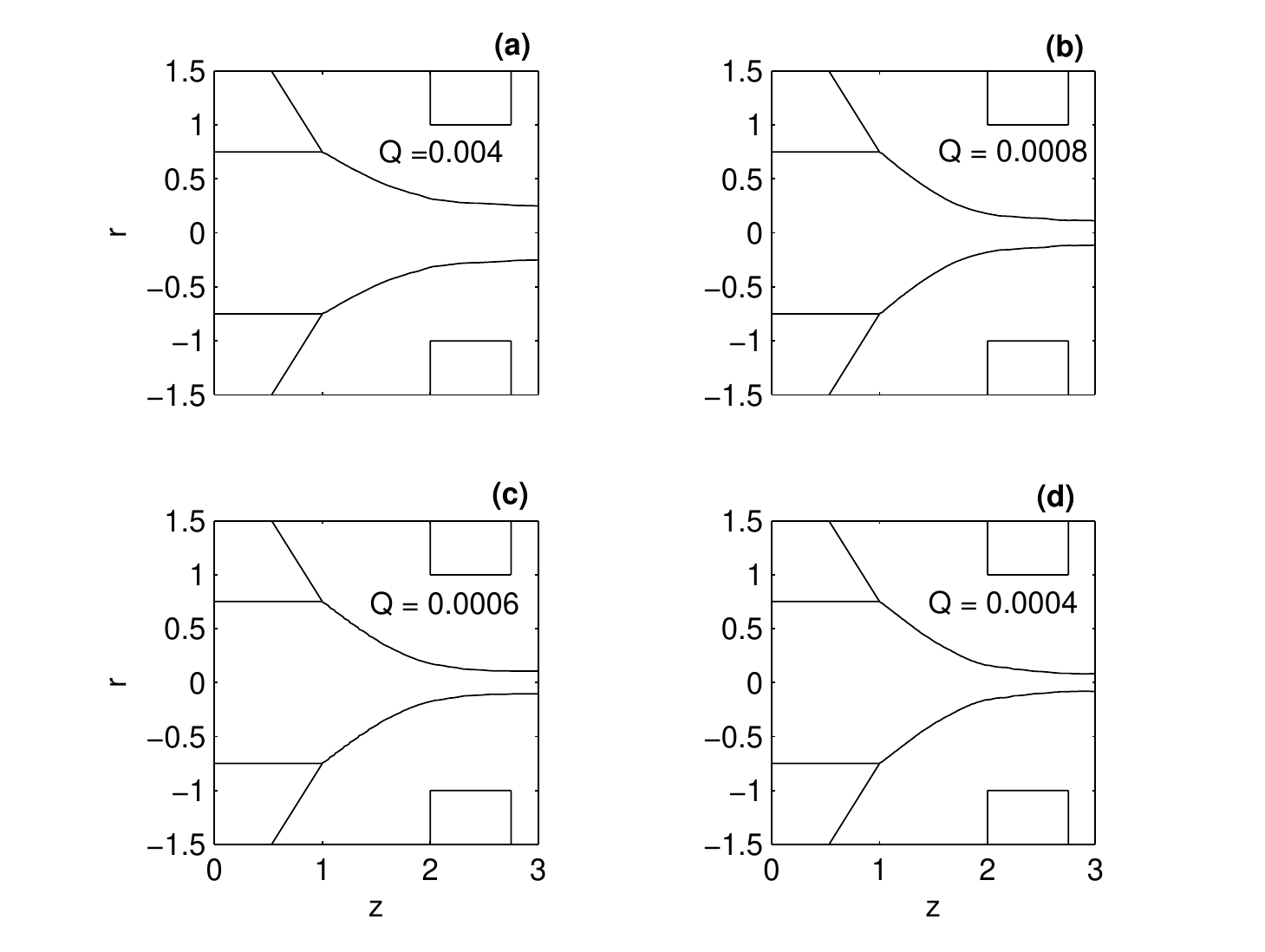}}
\caption{Shape of the liquid-gas interface as a function of $Q$
(case 1) } \label{f2-5}
\end{figure}

The procedure is the same for the two cases under consideration. The simulation is started from
rest with a value of $Q$ sufficiently high to obtain a steady jet. Then, $Q$ is reduced and the
solution is monitored in time until a new steady jet is obtained. Fig. \ref{f2-5} shows the
stabilized liquid-gas interface for case 1 and different $Q$. It should be pointed out that
$Q=0.0004$ is the smallest flow rate compatible with a steady jet for case 1. Therefore, it can be
identified as the minimum flow rate $Q^*$ for steady jetting: $Q^*=0.0004$ for case 1.

\begin{figure}
\centerline{\includegraphics[width=0.9\textwidth]{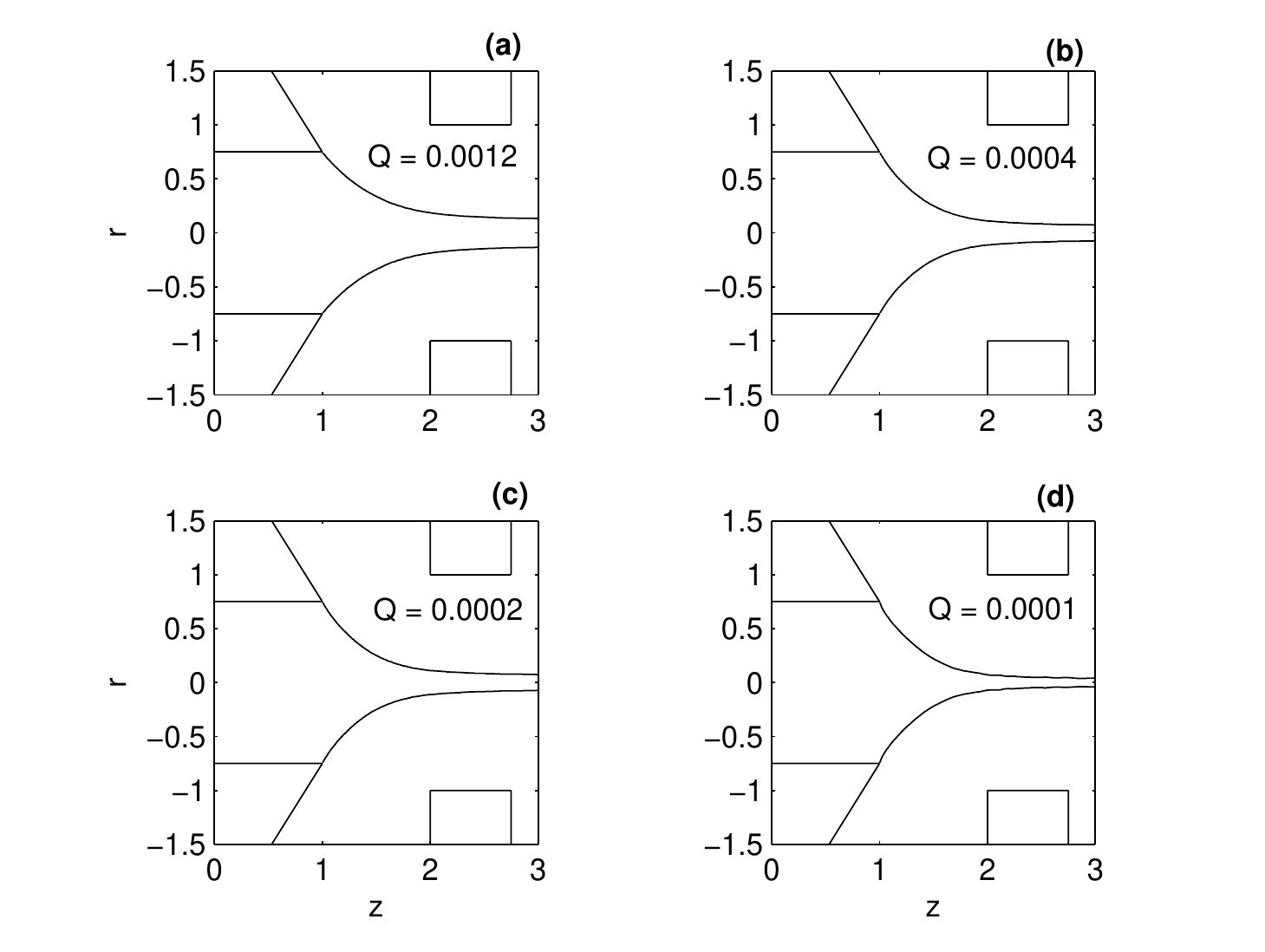}}
\caption{Shape of the liquid-gas interface as a function of $Q$
(case 2)} \label{f2-6}
\end{figure}

\begin{figure}
\centerline{\includegraphics[width=0.9\textwidth]{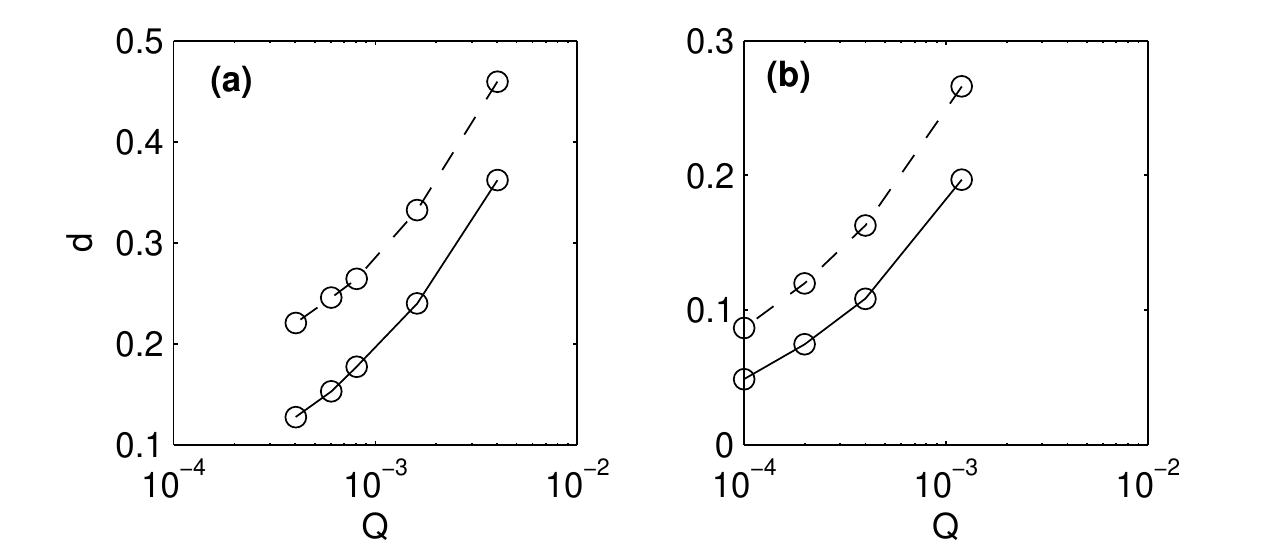}} \caption{Mean jet diameter at the inlet
(dashed line) and at the exit (solid line) of the nozzle orifice versus the flow rate quotient $Q$.
(a) Case 1, (b) Case 2.} \label{f2-7}
\end{figure}

\begin{figure}
\centerline{\includegraphics[width=0.9\textwidth]{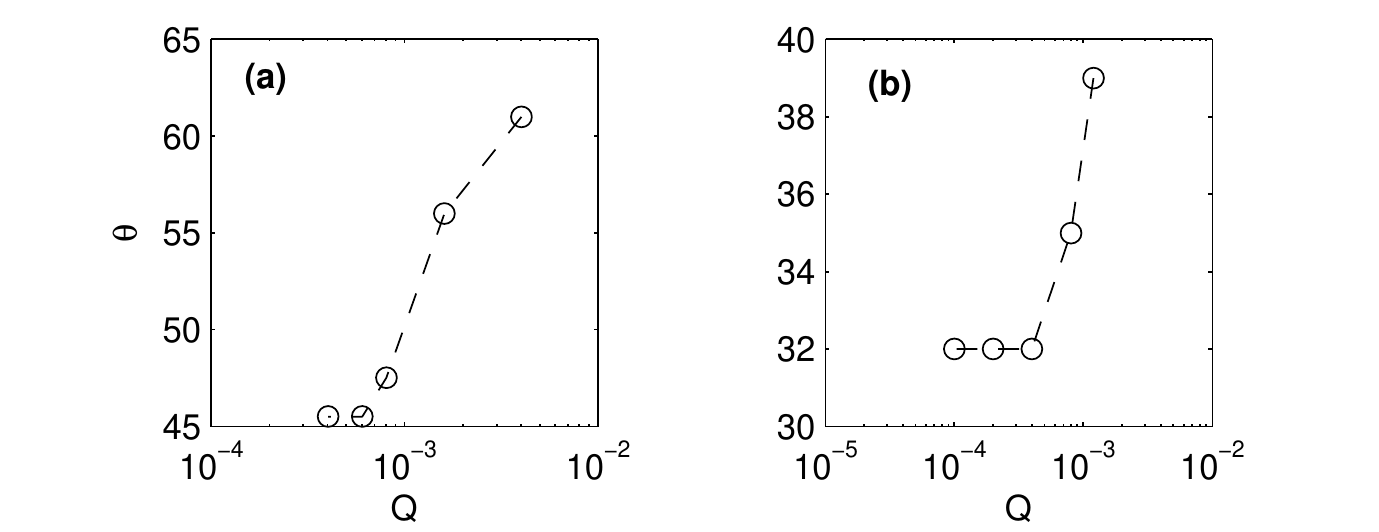}} \caption{Angle $\theta$ (degrees) versus
flow rate quotient $Q$. (a) Case 1, (b) Case 2.} \label{f2-8}
\end{figure}

Fig. \ref{f2-6} shows the interface for case 2 and different $Q$ values once a steady regime is
reached. The smallest jetting flow rate here is $Q^*=0.0001$, four times smaller than in case 1.
Accordingly, the smallest jet diameters are obtained for case 2. The jet diameter evolution is
shown in Fig. \ref{f2-7}, where the mean steady values $\bar{d}_{in}$ and $\bar{d}_{in}$ are
plotted as a function of $Q$ for (a) case 1 and (b) case 2. To complete the picture, Fig.
\ref{f2-8} shows the dependence of the meniscus angle $\theta$ with $Q$ for the two cases. In both
examples, $\theta$ becomes smaller as $Q$ decreases (smaller flow rate quotient implies stronger
focusing action). Just before dripping, as the liquid flow rate is reduced, the angle appears to
become independent from $Q$: the interface geometry becomes invariant (local hydrostatic balance).
The smallest angles are obtained in case 2. This is to be expected since the normal pressure forces
produced by the gas stream, which cause the focusing flow, are larger for that case.

\begin{figure}
\centerline{\includegraphics[width=0.9\textwidth]{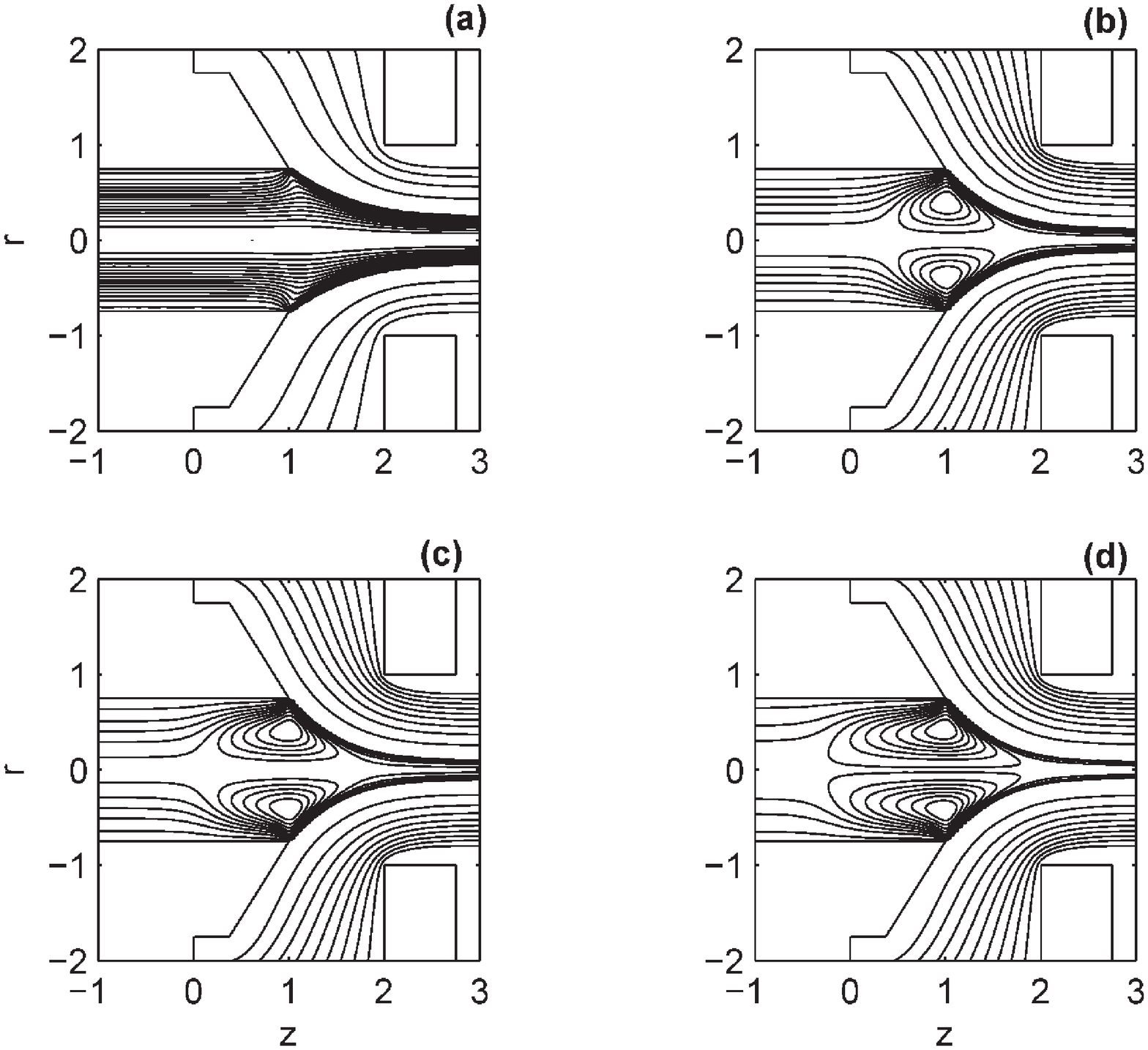}} \caption{Instantaneous streamlines at
four different flow rates quotients $Q$ for case 1:
 (a) $Q = 0.004$, (b) $Q = 0.0008$, (c) $Q = 0.0006$ and (d) $Q =
0.0004$.} \label{f2-9}
\end{figure}

\begin{figure}
\centerline{\includegraphics[width=0.9\textwidth]{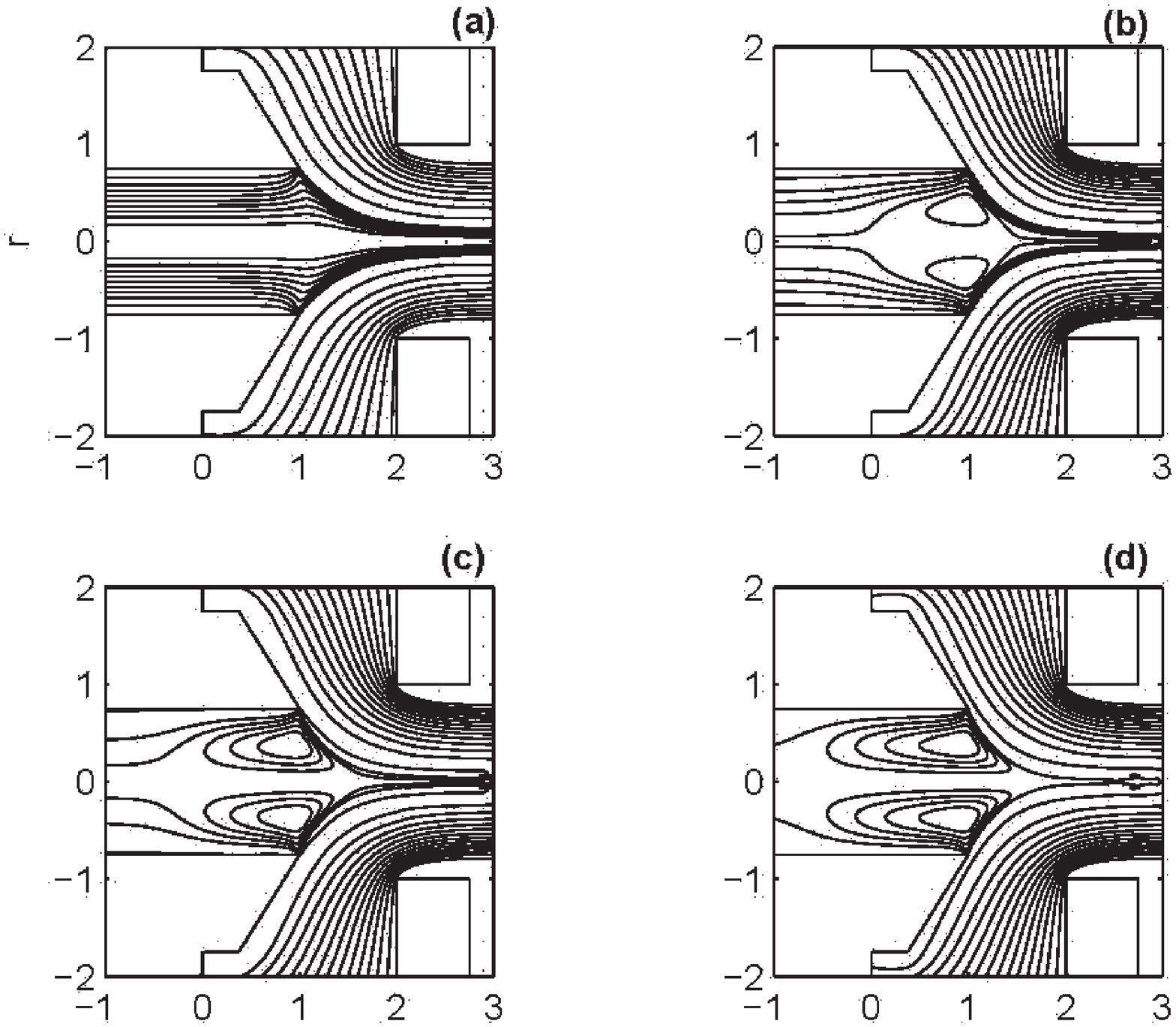}} \caption{Instantaneous streamlines with
four different flow rate quotients $Q$ for case 2: (a) $Q = 0.0012$, (b) $Q = 0.0004$, (c) $Q =
0.0002$ and (d) $Q = 0.0001$.} \label{f2-10}
\end{figure}

Analyzing in more detail the structure of the flow inside the liquid meniscus in jetting mode, in
the lower-$Q$ range, a meniscus recirculation cell is observed, in analogy with other co-flowing
systems \cite{SB2006,AnnaMayer06} and Taylor cones \cite{BG98}. Fig. \ref{f2-9} shows instantaneous
streamlines for case 1 and different values of $Q$. The recirculation increases when $Q$ decreases,
the cell penetrating into the capillary needle. Fig. \ref{f2-10} depicts instantaneous streamlines
for case 2 and four different values of $Q$. Again, a recirculation region appears before the
meniscus jet system ceases to be steady. The size of the recirculation can be calculated by finding
the two $z$-positions where the velocity at the axis becomes zero.

\begin{figure}
\centerline{\includegraphics[width=0.9\textwidth]{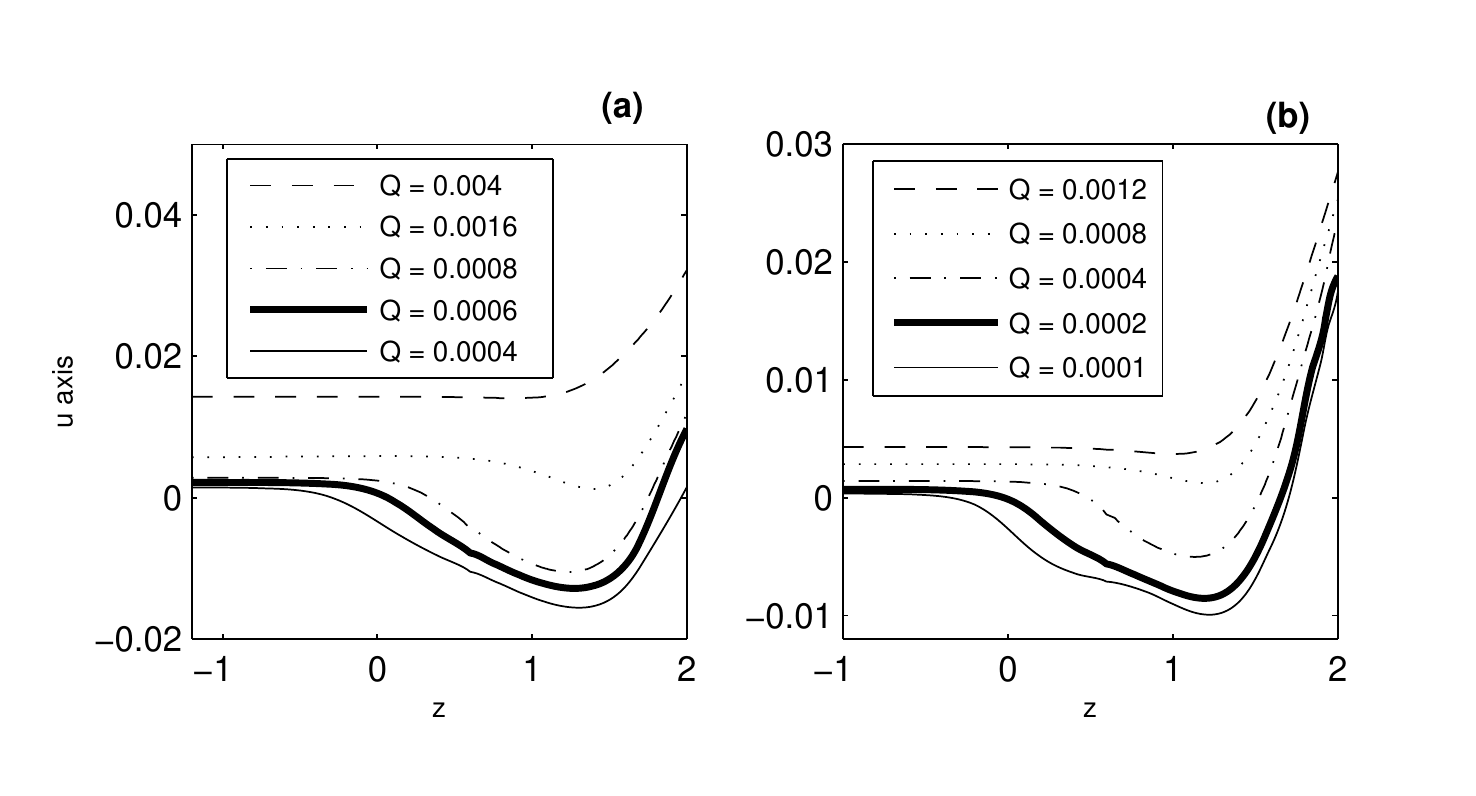}} \caption{Velocity at the axis versus
$z$ for several flow rate quotients $Q$: (a) Case 1 and (b) Case 2.} \label{f2-11}
\end{figure}

Fig. \ref{f2-11} shows the velocity at the axis, $v_{axis}$, as a
function of $z$ for five different values of $Q$: (a) case 1 and
(b) case 2. It is worth observing that $v_{axis}$ is roughly
uniform inside the capillary needle, its value being given by the
Hagen-Poiseuille expression; in the nozzle region it increases
owing to the focusing effect of the gas stream, which creates the
issuing jet. A region where $u_{axis}$ decreases is located in the
meniscus region, between the capillary and the nozzle; note that
when $Q$ decreases, a local minimum of $u_{axis}$ is observed at a
given position $z=z_{min}$ in the meniscus region. If $Q$ is
sufficiently small, $v_{axis}$ becomes negative near the local
minimum in a region delimited by the two $z$ positions, $z_1$ and
$z_2$, where $v_{axis}=0$. Therefore, the size of the
recirculation region, $S_r$, observed in Figs. \ref{f2-10} and
\ref{f2-11}, can be computed as $S_r/R=s_r=z_2-z_1$. There is a
threshold value of $Q$, $Q_r$, below which a recirculation pattern
is observed. At the threshold flow rate $v_{axis}=0$ at
$z_1=z_{min}=z_2$ and $v_{axis} > 0$ elsewhere.

\begin{figure}
\centerline{\includegraphics[width=0.9\textwidth]{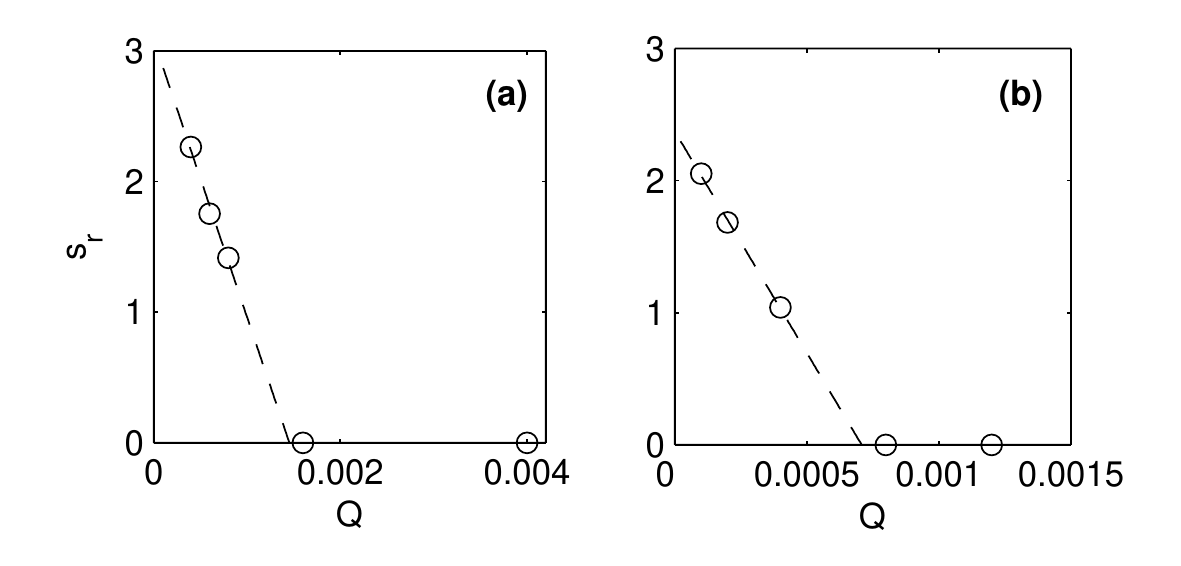}} \caption{Size of the recirculation
$s_{r}$ as a function of $Q$: (a) Case 1 and (b) Case 2. } \label{f2-12}
\end{figure}

Fig. \ref{f2-12} shows $s_r$ as function of $Q$ for (a) case 1 and
(b) case 2. Looking back at Fig. \ref{f2-9}, note that the size of
the recirculation cell increases as $Q$ decreases. In situations
of incipient recirculation ($Q$ smaller than but similar to $Q_r$)
this growth appears to be linear, as derived later from
dimensional arguments. In Fig. \ref{f2-12}, the discrete points
'o' have been obtained directly from the simulations. The dashed
lines are linear interpolations computed in the recirculating
regime, $s_r>0$. The linear interpolation is not only in good
agreement with the data but also provides a reliable approximation
to compute $Q_r$. The estimations are $Q_r=0.001453$ for case 1
and $Q_r=0.000708$ for case 2. According to the above, the linear
expression relating the size of the recirculation region and the
flow rate $Q$ is: \be s_r\sim A (Q_r-Q). \label{RL}\ee This means
that $s_r$ is proportional to a \it back flow \rm rate
$Q_B/Q_g=Q_b=(Q_r-Q)$ for a given set of fluid properties,
geometry, and gas flow Reynolds number. The relative location of
the jetting threshold $Q^*$ and the recirculation threshold $Q_r$
is, in both cases, $Q^*<Q_r$. Therefore, recirculation can be
taken as a dripping-precursor: as far as can be gathered from our
simulation, it always precedes global instability of the
meniscus-jet system.

\begin{figure}
\centerline{\includegraphics[width=0.9\textwidth]{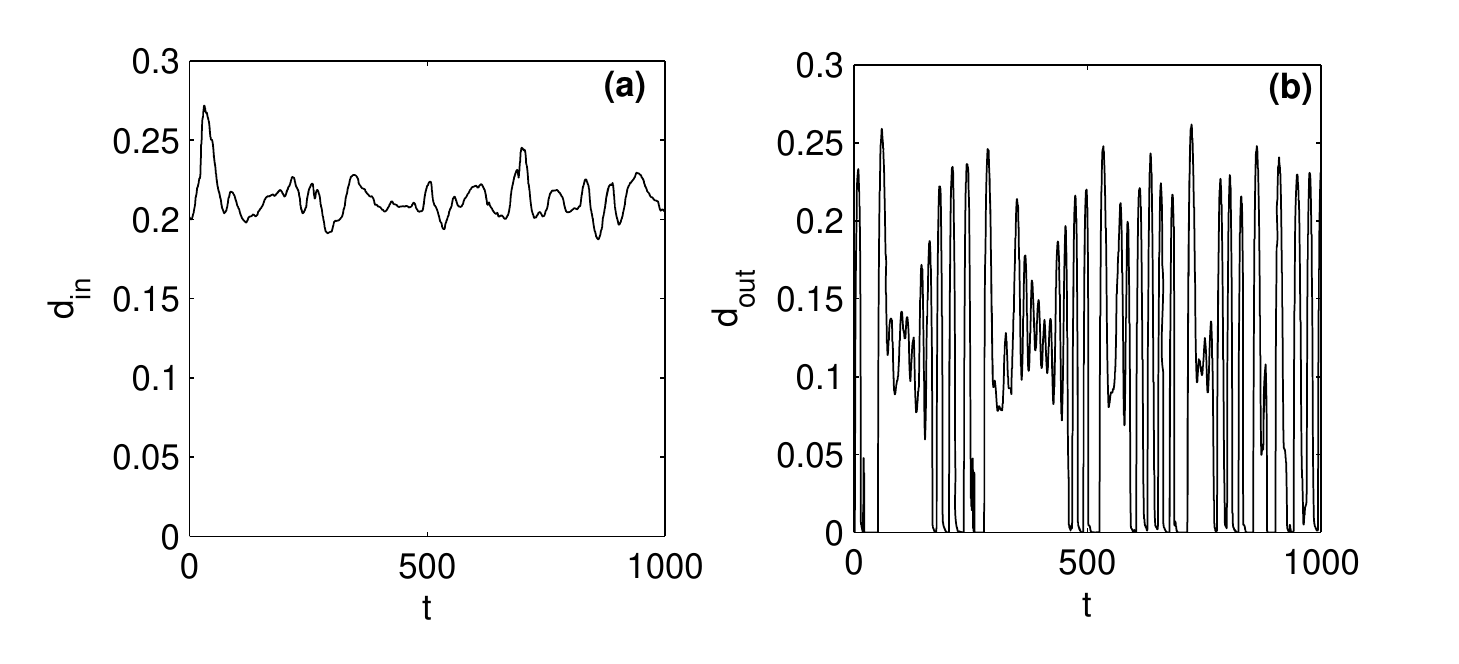}}
\caption{Time evolution of (a) $d_{in}$ and (b) $d_{out}$ in a
irregular dripping regime for case 1 and
 $Q=0.000322
$.} \label{f2-14b}
\end{figure}

\begin{figure}
\centerline{\includegraphics[width=0.9\textwidth]{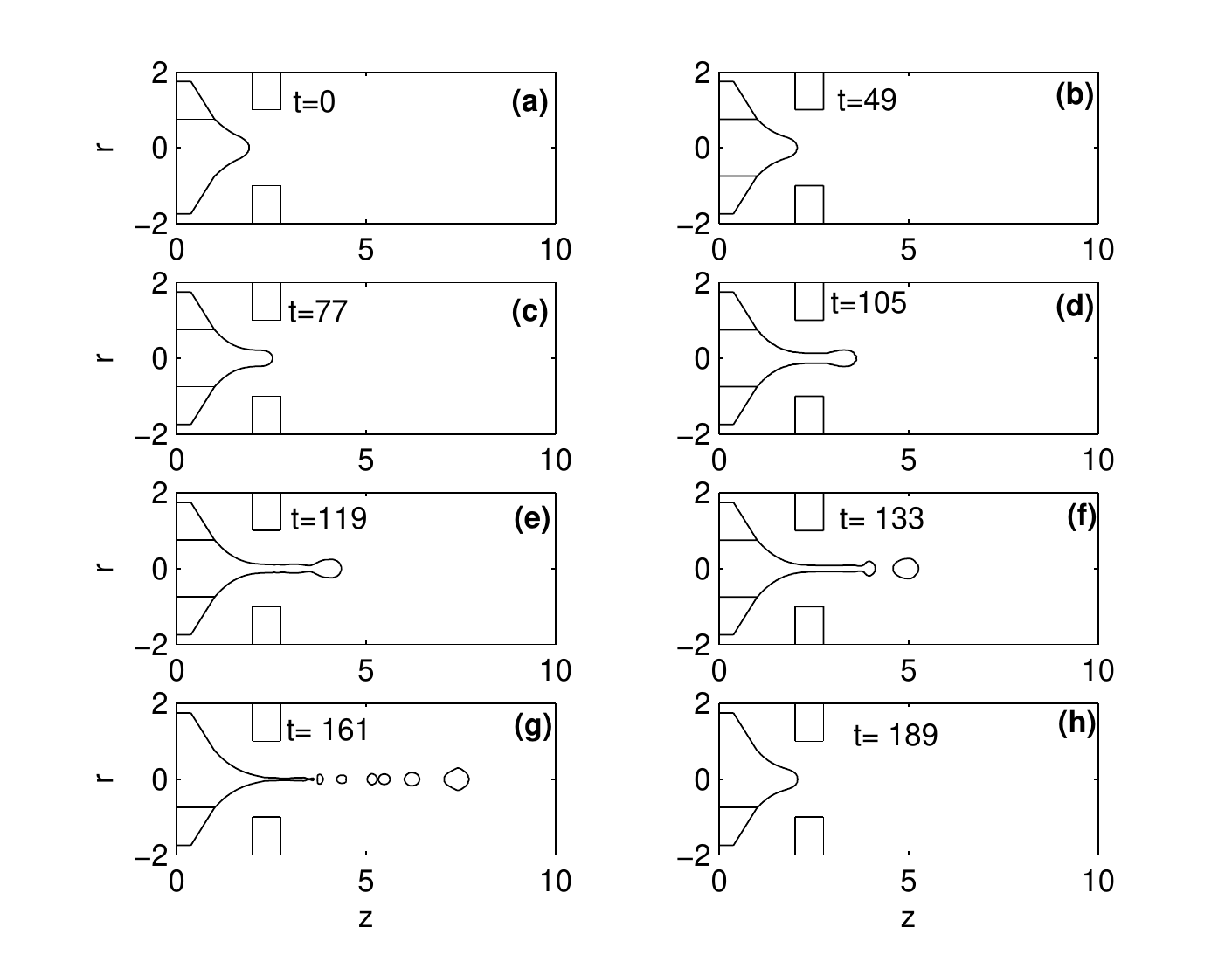}} \caption{A sequence of the dripping
regime for case 1 and $Q=0.000241$ involving a complete cycle.} \label{f2-13}
\end{figure}

\begin{figure}
\centerline{\includegraphics[width=0.9\textwidth]{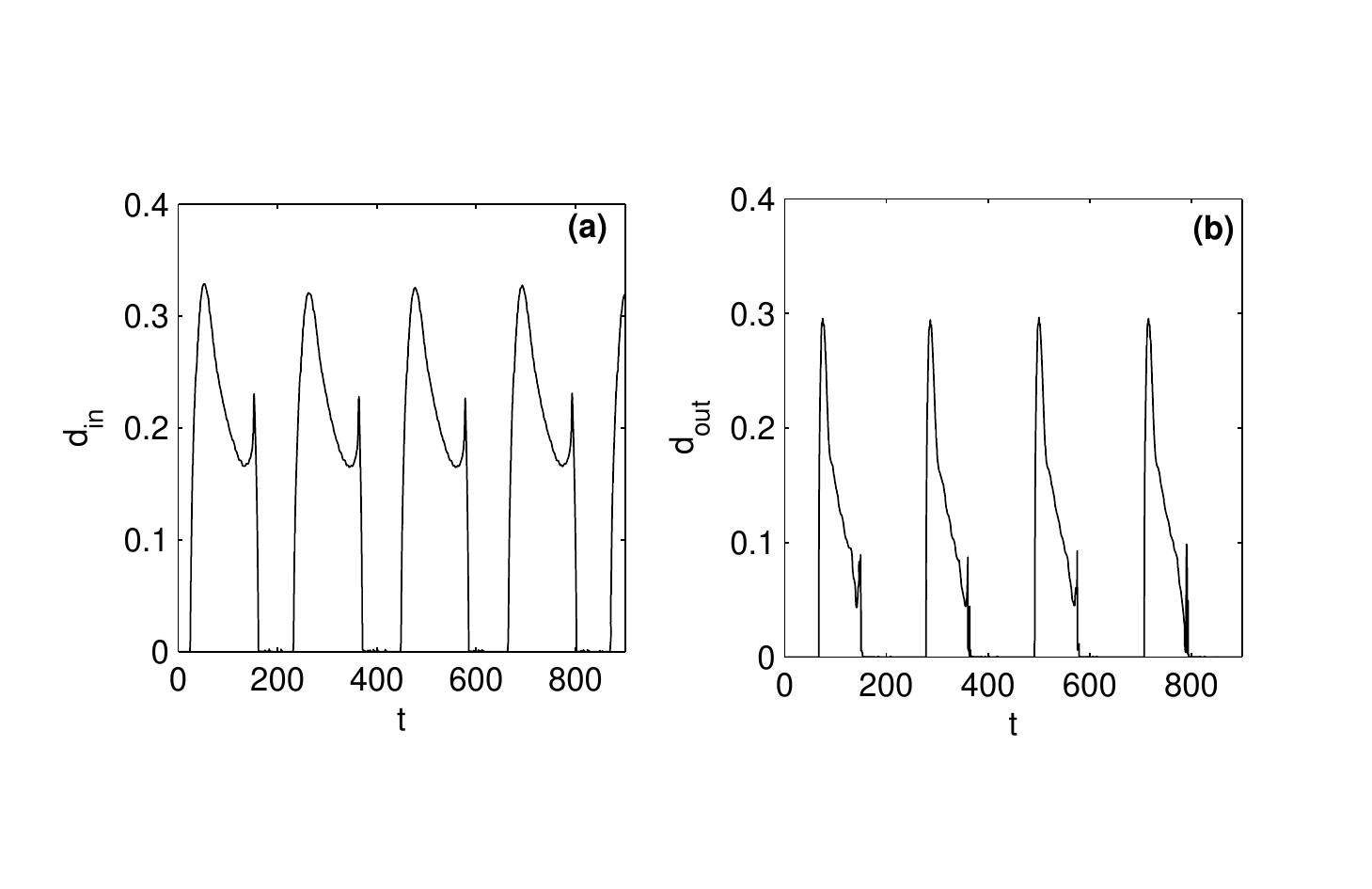}} \caption{Time evolution of (a) $d_{in}$
and (b) $d_{in}$ in a dripping regime for case 1 and $Q=0.000241 $.} \label{f2-14}
\end{figure}

Finally, some results are presented with flow rate values below the jetting threshold.
For $Q<Q^*$, with $Q$ close to $Q^*$, our simulations show the flow to exhibit different
behaviors in a sequence: a period where a thin jet breaks up in the nozzle region
alternates with other periods where a thin jet breaks up downstream of the nozzle. The
irregular time behavior of the flow for $Q<Q^*$, but $Q$ \emph{close to} $Q^*$
(\emph{incipient dripping}) can be observed in Fig. \ref{f2-14b}, where $d_i$ (a) and
$d_{out}$ (b) are shown as a function of time for case 1 and $Q=0.000322$. For $Q<Q^*$
but $Q$ \emph{sufficiently different from} $Q^*$, the flow behavior becomes more regular
and periodic with a unique dripping frequency.

As anticipated at the first part of this section, our results,
while belonging to the dripping-regime, are untypical in that the
co-flowing current gives rise to axial stretching of the jet and
drops, so that unusual breakup geometries result. The radial
extension of the drops is limited, and the axial stretching gives
rise to secondary breakup, immediately after detachment. The
pattern observed in Fig. \ref{f2-13} points to a dripping regime:
it is perfectly periodic (transient jetting can therefore be
excluded) and each period is associated with the filling up of a
drop, its breakup from a thinning filament, and the recoil of this
filament. Fig. \ref{f2-13} shows a complete time sequence of a
dripping process (case 1, $Q=0.000241$). Subfigures (a)-(e) show
the growth of the meniscus and the formation of a jet issuing from
the nozzle; subfigures (g)-(h) show the jet breakup into droplets
of different sizes and the meniscus recoil. This is a periodic
sequence, the period being $T \sim 210$ for each cycle. Fig.
\ref{f2-14} plots $d_i$ (a) and $d_{out}$ (b) as a function of
time for this case. Initially, a liquid meniscus is growing with
no jet production, and $d_{in}=d_{out}=0$. Subsequently, a liquid
jet issues and $d_{in}$ and $d_{out}$ become positive. Both
diameters reach a maximum at a certain time and then start to
decrease. Finally, the jet breaks into droplets, $d_{in}$ and
$d_{out}$ are set to zero and the process begins anew. In spite of
the observed differences, the dripping process in this case is
quite similar to regular faucet dripping, the time period being
mainly imposed by the filling of the meniscus until reaching a
critical volume.

\begin{figure}
\centerline{\includegraphics[width=0.9\textwidth]{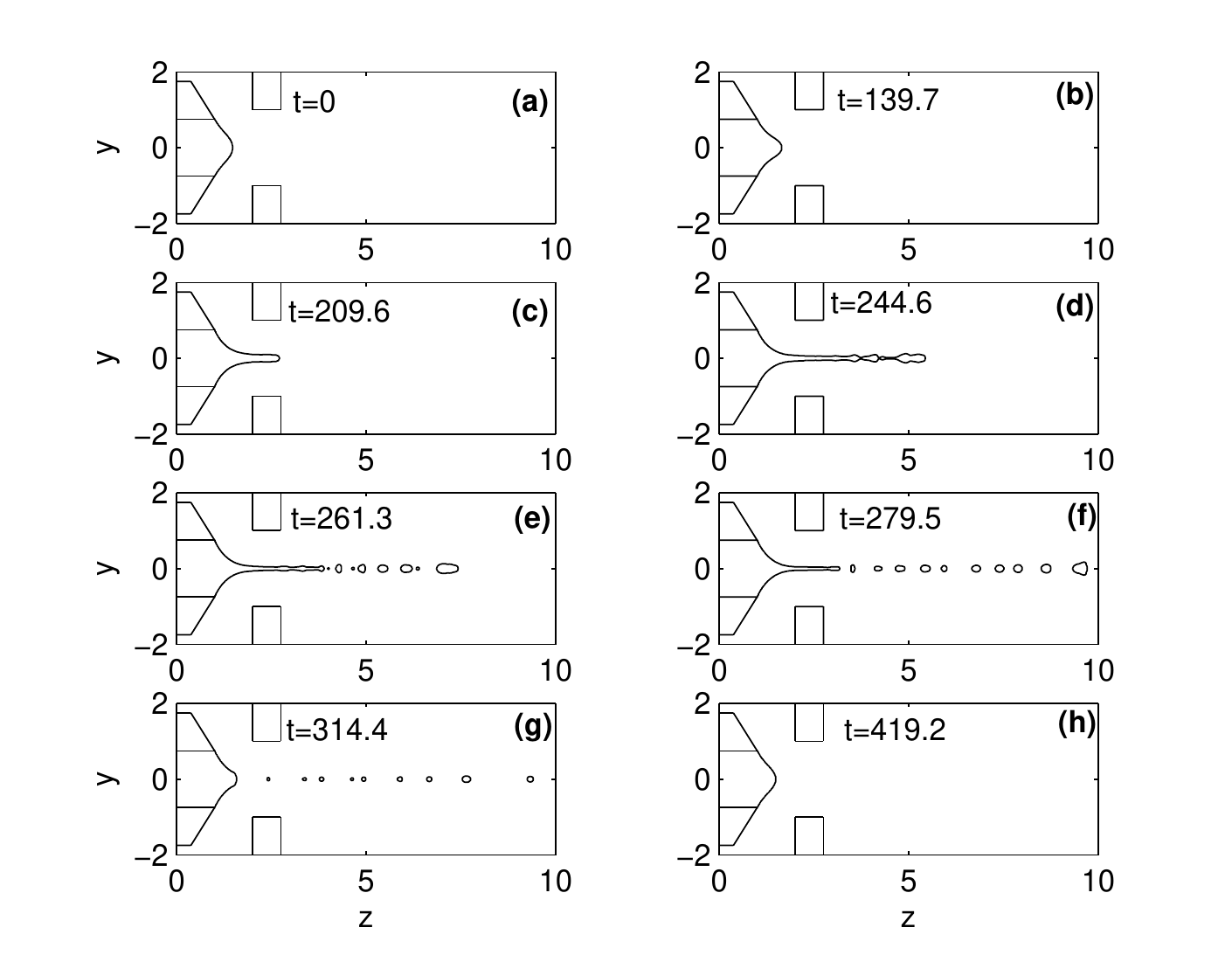}} \caption{A sequence of the dripping
regime for case 2 and $Q=0.00004$ involving a complete cycle.} \label{f2-15}
\end{figure}

\begin{figure}
\centerline{\includegraphics[width=0.9\textwidth]{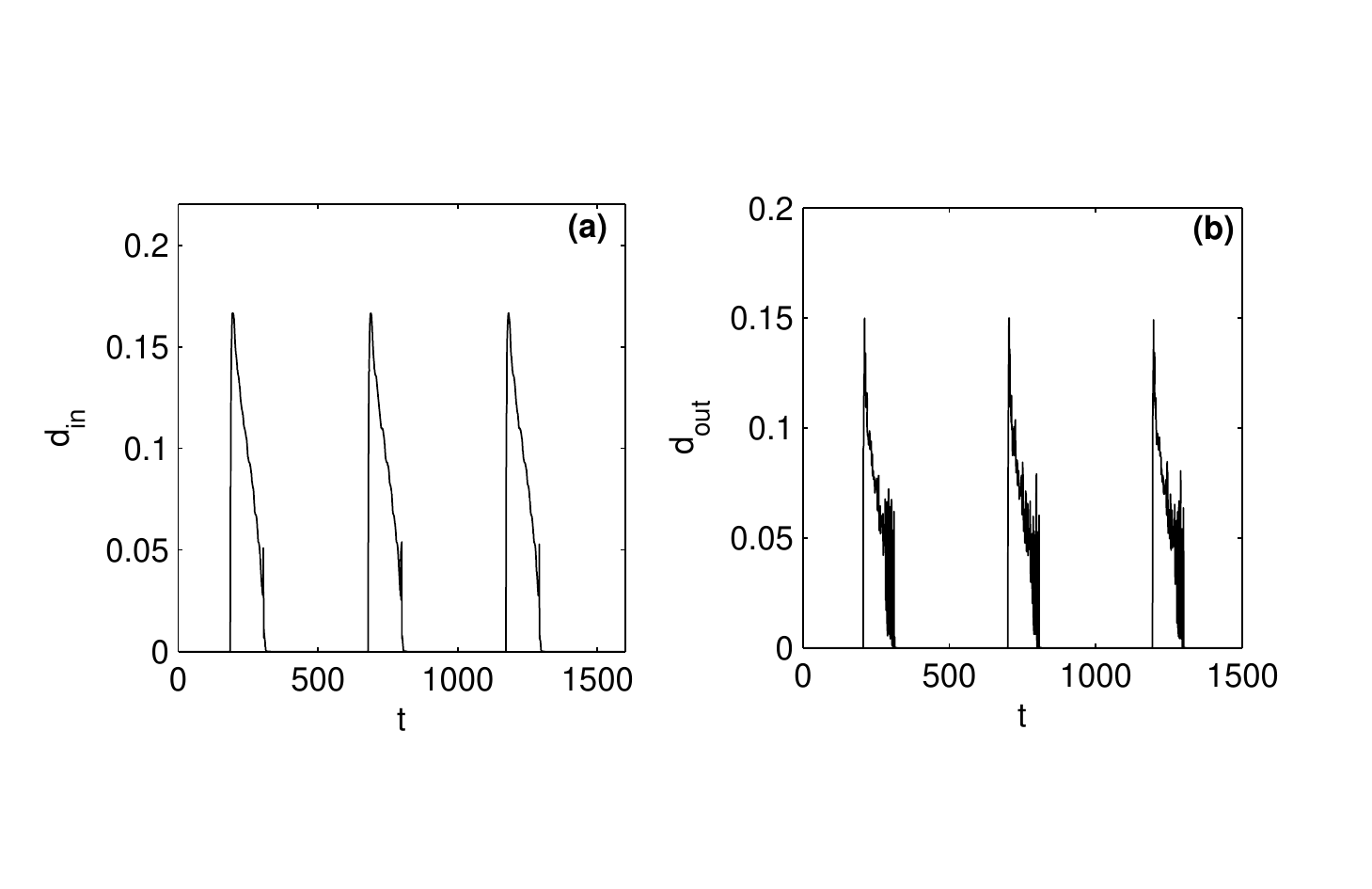}}\caption{Time evolution of (a) $d_{in}$
and (b) $d_{in}$ in a dripping regime for case 2 and $Q=0.00004$} \label{f2-16}
\end{figure}

A similar situation is observed in case 2. Fig. \ref{f2-15} shows
a complete time sequence of a regular dripping process with
$Q=0.00004 $. Subfigures (a)-(d) show the meniscus growth and the
emission of a jet, much thicker and longer than observed in Fig.
\ref{f2-13}. Again, this sequence is time periodic with a period
$T\sim 500$  for each cycle (see $d_i$ (a) and $d_{out}$ (b) in
Fig. \ref{f2-16} as a function of time).

\subsection{Influence of the BCs and the spatial and temporal
resolution on the numerical results}

The numerical problem addressed is quite complex: it involves a
high speed stream of gas discharging through a nozzle into a
infinity large chamber plus a meniscus-liquid jet which may break
into droplets within the finite numerical domain. This complexity
leads to different time and spacial scales associated to a
plurality of interacting physical phenomena (jet breakup due to
capillary and Kelvin-Helmholtz instabilities, mixing layer
instabilities in the main gas stream). Therefore, an accurate
analysis of the jet breakup is difficult to achieve. On account of
it, though our VOF method is fully reliable in qualitative terms
as a predictor of jet breakup and drop formation, we have devoted
this subsection to check that our numerical results (meniscus-jet
shape as a function of $Q$ for different setups) are independent
of the selected BCs and numerical meshes.

\begin{figure}
\centerline{\includegraphics[width=0.9\textwidth]{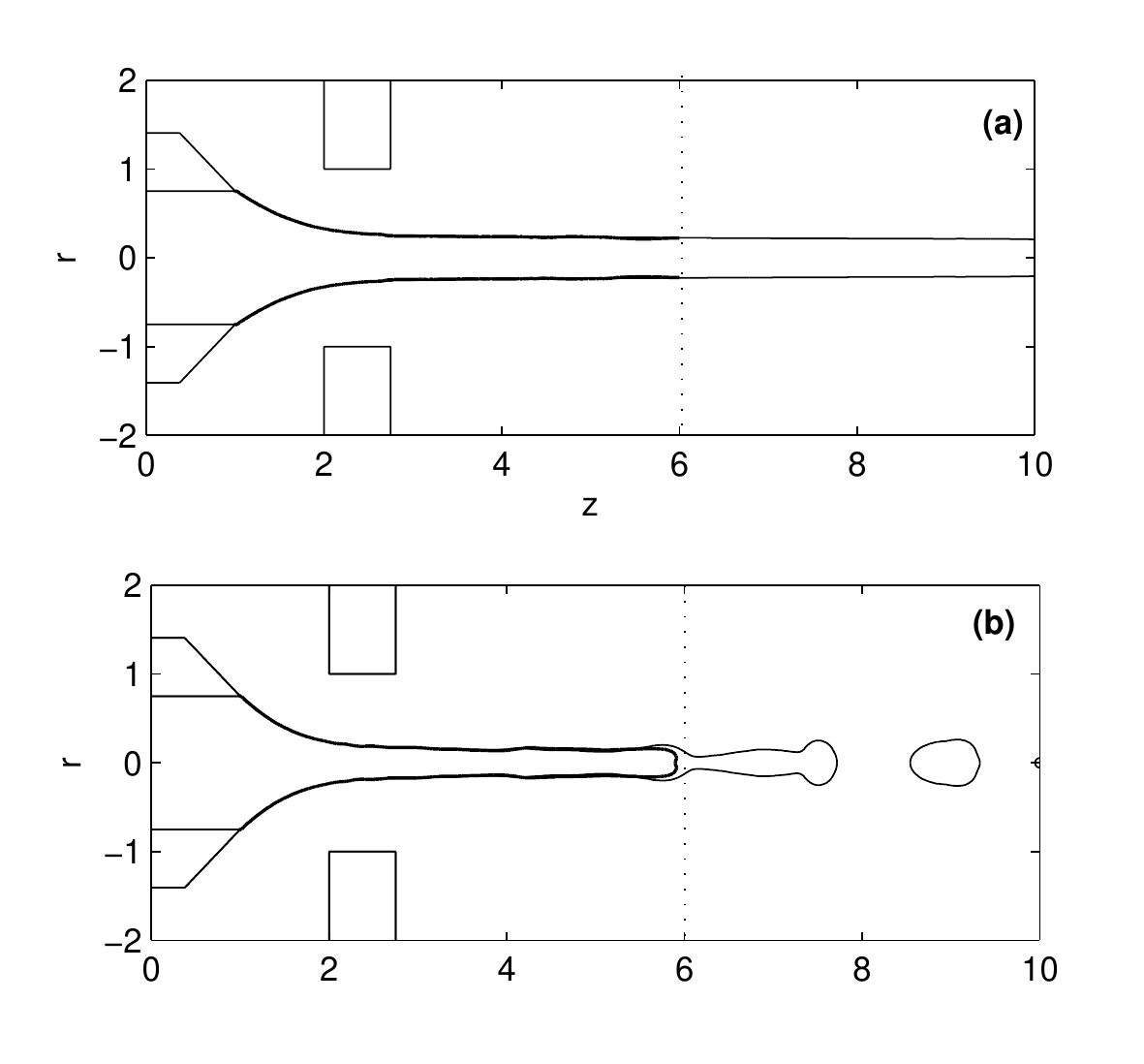}}
\caption{The effect of the $p=0$ boundary condition on the
liquid-gas interface comparing to downstream boundary locations:
$z=6$ (thicker line) and $z=10$ (thinner line):(a) data as in case
1 and $Q=0.004 $; (b) data as in case 1 and $Q=0.0008$.
}\label{f2-16new}
\end{figure}

As indicated above, the most problematic simulation choice is
setting $p=0$ at the outlet boundaries, since any jet or a drop
crossing the boundary is influenced by the strong and artificial
restriction that the pressure remains fixed. Our choice is a
simplification ($p = 0$) which takes advantage of the essentially
parabolic character of the equations. A second option has been
explored, the so-called outflow conditions (assuming uniformity,
i.e. Neumann type), but they give rise to a false constraint on
the flow pattern, because they imply that the gas flow is coaxial.
A minisymposium held in 1994 on the open boundary condition
problem in incompressible flow, by Sani and Gresho \cite{OBC} led
to the concluding remark: ``We have made some attempts at shedding
more light on the difficult and unresolved area of seeking good
OBCs for incompressible flow simulations. It has been an exercise
in frustration and we are not thrilled with the results
obtained''.

There is an evident inaccuracy involved in our $p=0$ choice: the
pressure jump associated to an interface will lead to high local
pressure inside the liquid jet or droplets. However, this
assumption can be reconciled with our aim, which is not a study of
the breakup process and its transient geometries. We are
addressing a wider scale: the cone-jet flow pattern, and the
general drop generation regime. To show that the distortion caused
by this artificial BC is local and does not modify the global
behavior at the cone-jet region, some exploration as been carried
out. It can be shown that setting the external boundary
sufficiently far downstream from the nozzle region, at
$z_{out}=10$, the meniscus-jet is not affected by the boundary
condition. To show this, we have considered the worst scenario: we
choose large liquid flow rates $Q$ and weak gas flow (case 1).
Fig. \ref{f2-16new} shows the stabilized liquid-gas interface for
case 1 and two different values of $Q$, computed in the original
domain and in a shorter one. In Fig. \ref{f2-16new}(a), the jet
does not break up within any of the two numerical domains and the
jet and meniscus interface in the nozzle region is evidently not
affected by the artificial $p=0$ boundary condition. The influence
of the artificial BC is confined to a few diameters upstream of
the downstream boundary. In the case considered in Fig.
\ref{f2-16new}(b), the jet is breaking up into drops within the
large domain. Even in this case, the meniscus and jet interface in
the nozzle region is not affected by the artificial boundary
condition.

\begin{figure}
\centerline{\includegraphics[width=0.9\textwidth]{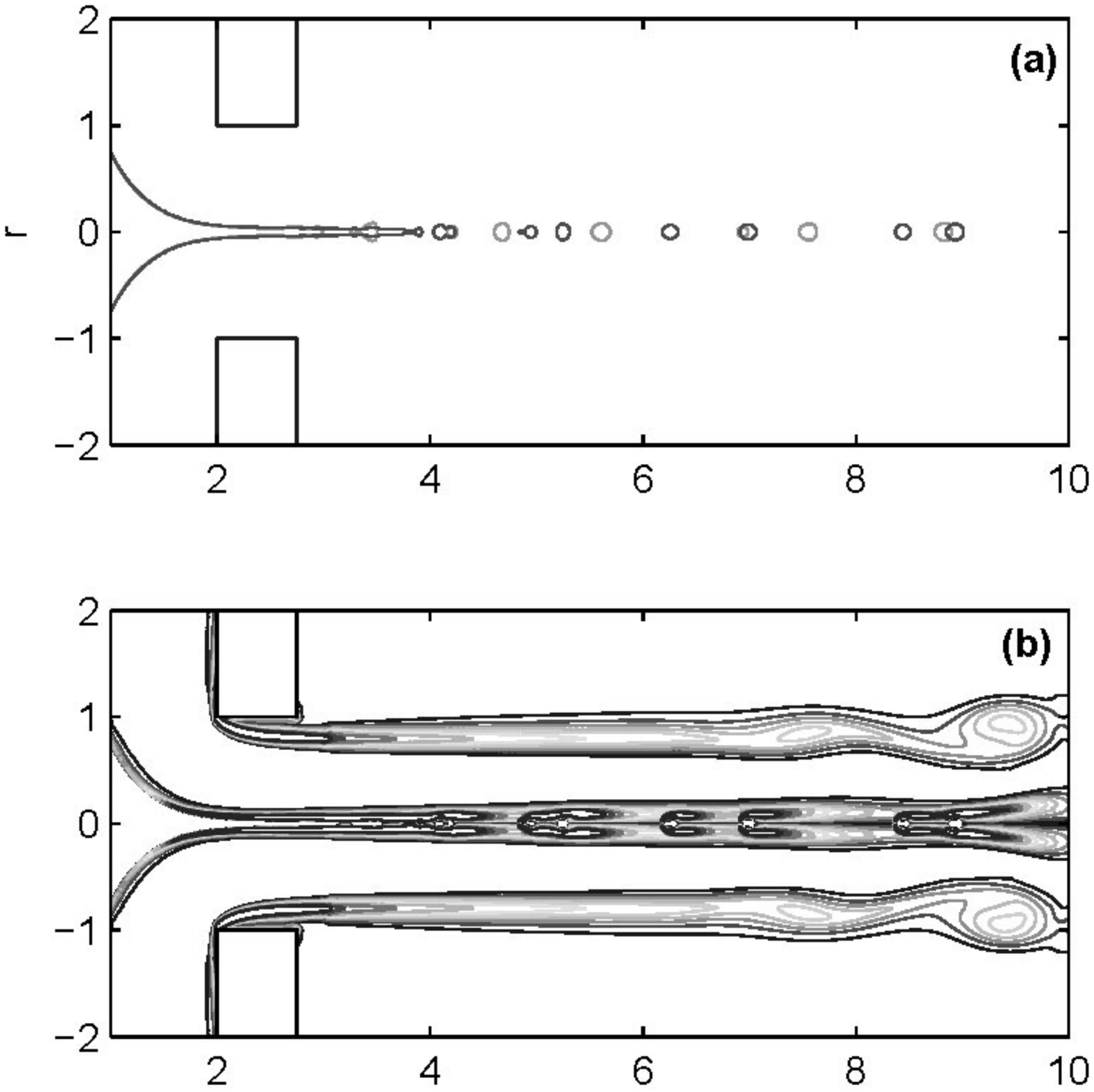}}
\caption{The effect on the liquid-gas interface of the spatial
resolution: (a) a general view of case 2, $Q=0.0001$ where the
grey line is the solution computed in a mesh where
$(\Delta_{z})_{min}=(\Delta_r)_{min}=0.01$ and the black line is
the solution obtained in a mesh with
$(\Delta_{z})_{min}=(\Delta_r)_{min}=0.005$; (b) contours of the
vorticity field for the same case with the best spatial
resolution.}\label{f2-16new1}
\end{figure}

\begin{figure}
\centerline{\includegraphics[width=0.9\textwidth]{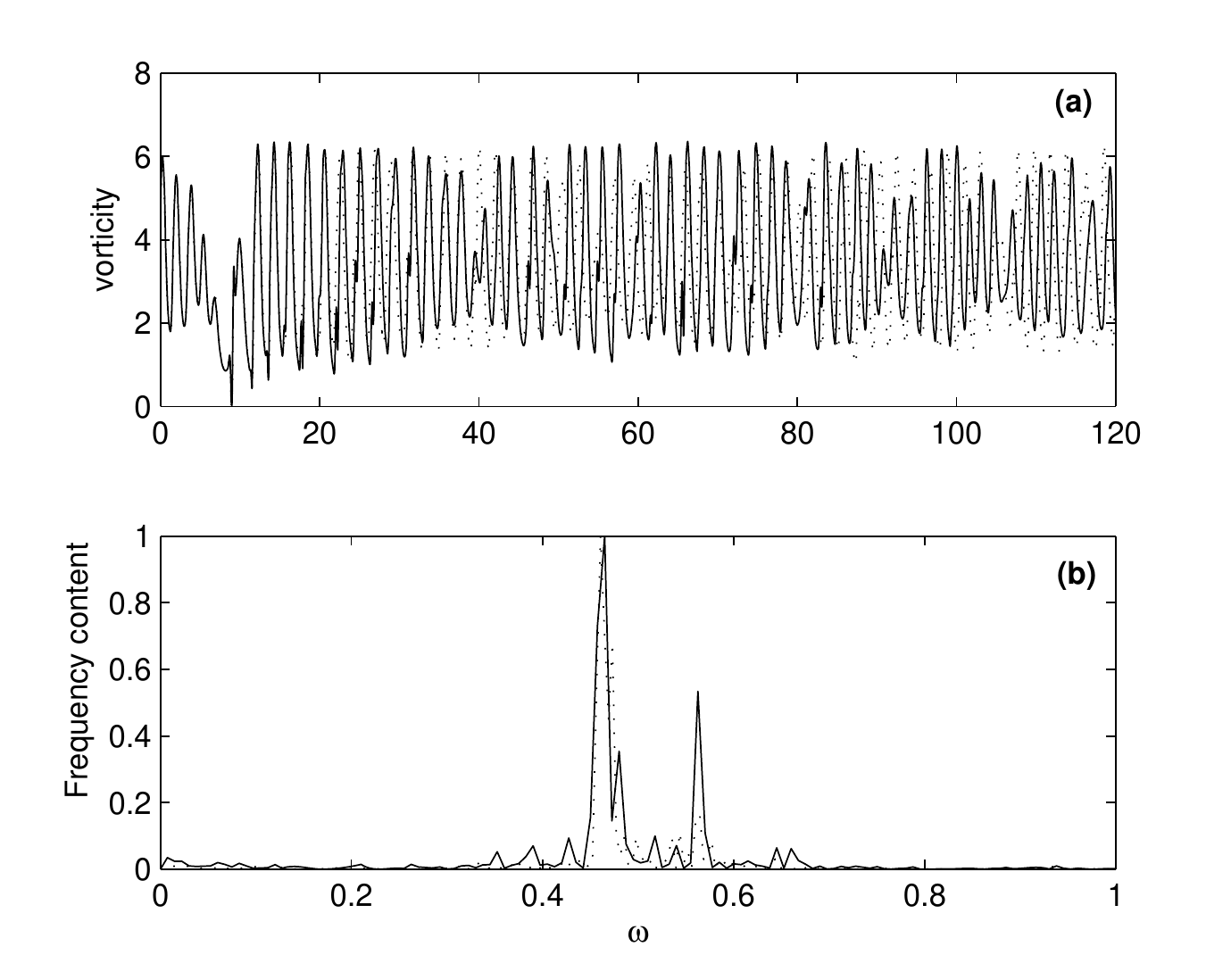}}
\caption{ a) Comparison of the time evolution of the vorticity
magnitude at the point ($z=9.5$, $r=0.995$) for case  2 and
$Q=0.0001$ computed with $\Delta t=0.014$ (solid line) and with
$\Delta t=0.028$ (dashed line); b) shows the frequency content of
the two signals.}\label{f2-16new2}
\end{figure}

Let us now show the consistency of the model by comparing the results in two different meshes. In
this case again, we have considered the worst scenario, by selecting smaller values of the liquid
flow rate ($Q$ small) and a large gas flow (case 2), since thinner jets are obtained in these
cases. Fig. \ref{f2-16new1} (a) shows a instantaneous picture of a steady (convectively unstable)
liquid jet breaking up into drops (case 1, $Q=0.0001$) computed in two different meshes. Observe
that although we are comparing the liquid interface at two different times and with different
spatial resolution, the shape of the meniscus of the liquid in the nozzle zone coincides. The main
difference is that the liquid jet is slightly longer in the finer mesh. As mentioned, an accurate
description of the jet breakup is not the objective of this paper. It requires specific analytic
tools in order to capture the diverse physical phenomena involved.

This can be illustrated by Fig. \ref{f2-16new1} (a), showing
instantaneous contours of the vorticity field for this case,
computed with the best spatial resolution. It can be seen that,
owing to the large density and velocity difference between the
liquid and the gas, the vorticity is large in the liquid-gas
boundary layer which develops at the liquid-jet meniscus. When the
jet breaks up, the behavior of the flow around the drops is
similar to high Reynolds number flow around a round object (since
the gas is travelling faster that the liquid drops). Therefore,
boundary layer separation is to be expected at the surface of the
drops, and a wake will originate downstream of the drop. In
addition to this, a mixing layer is developing downstream of the
nozzle edge owing to the difference between the velocity of the
gas flowing through the nozzle and the stagnant atmosphere
surrounding the outlet. This mixing layer yields to the
development of vortices as shown in the figure. Fig.
\ref{f2-16new2}(a) shows the time evolution of the vorticity
magnitude at a point of this mixing layer computed with two
different time steps. Strong fluctuations of the vorticity
magnitude are apparent. The flow evolution shows some sensitivity
to the resolution, particularly as the simulation time increases.
However, both time resolutions are reliable to predict the
characteristic frequencies of the problem at the observation
point. Fig. \ref{f2-16new2}(b) show the frequency content of the
vorticity as obtained by applying the Fast Fourier transform (FFT)
to the two signals. The main frequency $\omega_1\sim 0.46$ is
related to the passage of the vortices generated in the gas mixing
layer. There is a secondary characteristic frequency,
$\omega_2\sim 0.56$, associated to the interaction between the
mixing layer and the vorticity wake of the drops. This
interpretation was strengthened by recomputing the flow without
the liquid jet: the frequency content of the vorticity signal at
the same observation point only showed the main frequency peak
$\omega_1\sim 0.46$ .

\section{Comparison with analytical models and scaling laws}

The first predictive model for the jet diameter $d_j$ at the
orifice exit \cite{Gan98} assumes that viscous and capillary
effects are small enough compared to liquid inertia. This demands
large enough Reynolds and Weber numbers of the liquid jet, in
reasonable agreement with most experimental conditions (common
solvents including water, down to the micron scale). In this
limit, the overall pressure difference $\Delta P=\Delta P (Q_g)$
(pressure difference between the gas inlet and the gas outlet)
imposed in the downstream direction (i.e. through the orifice),
transmitted to the liquid stream by normal surface stresses, is
converted into kinetic energy, so that \be \Delta P \simeq
\fr{1}{2} \rho_l v^2\simeq \fr{8 Q_l^2}{\pi^2 d_j^4},\ee which
readily gives \be d_j=\left(\fr{8\rho_l}{\pi^2\Delta
P}\right)^{1/4} Q_l^{1/2}.\label{ff}\ee Furthermore, the jet is
assumed sufficiently small compared to the orifice diameter $D$
such that no only it does not touch the orifice borders, but also
the boundary layers of the focusing fluid (gas) at the orifice and
at the jet's surface are sufficiently small compared to the corona
defined between the jet and the orifice. This is why $D$ does not
enter expression (\ref{ff}). Neither does $D_{1}$ have any direct
influence on the jet diameter; only as a parameter determining the
liquid flow rate $Q_l$.

\begin{figure}
\centerline{\includegraphics[width=0.9\textwidth]{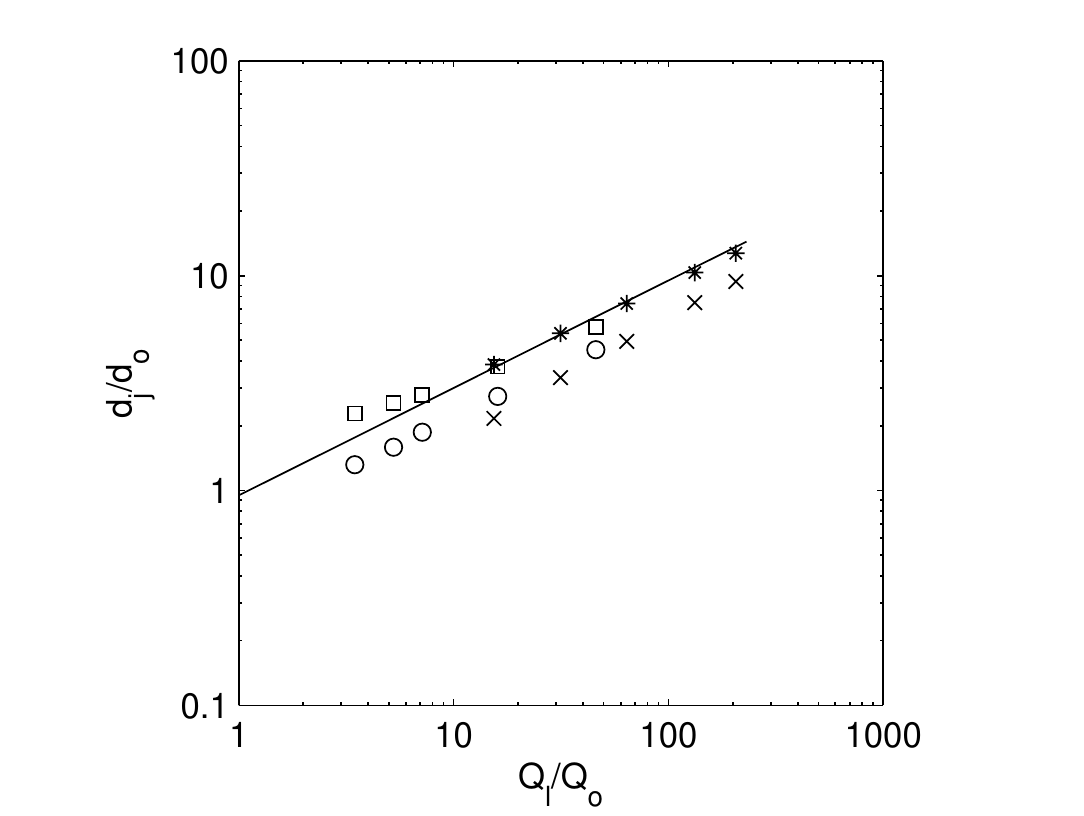}}
\caption{Jet size measured at the entrance of the nozzle using
$\bar{d}_{in}$ ($\Box$  correspond to case 1 and $*$ to case 2)
and at the exit using $\bar{d}_{out}$ ($o$  correspond to case 1
and $\times$ to case 2) compared to the theoretical prediction
(continuous line).}\label{f2-17}
\end{figure}

Interestingly, if viscous effects and surface tension are
neglected, and we assume $d_j<<D$, the only operating parameters
left in the analysis are $\{\rho_l,\Delta P, Q_l\}$; using these
three parameters, a scaling law identical to \ref{ff} -regardless
the constants- follows from dimensional reasoning. Figure
\ref{f2-17} illustrates the accuracy of this first simple
prediction. However, that expression does not provide information
on \it how small \rm the neglected effects are. Would it be
possible to quantify \it both \rm the dependence of the jet
diameter on the three main parameters and the relative magnitude
of each one of the neglected effects? The answer is yes, by
retaining either $\sigma$, $\mu_l$, or $D$ in the dimensional
analysis, and using $\{\Delta P,\sigma,\rho_l\}$, $\{\Delta
P,\mu_l,\rho_l\}$ or $\{\Delta P,D,\rho_l\}$ as independent
parameters, respectively.

In physical terms, the relative effect of surface tension may be
determined by observing that the liquid Weber number must be of
order unity (${\rm We_l}=\rho_l 8Q_l^2/(\pi^2\sigma d_j^3)$) for a
given pressure $\Delta P$, from which one obtains the limiting
diameter $d_o$ and flow rate $Q_o$. The result is
$d_o=\sigma/\Delta P$ and $Q_o=\left(\sigma^4 \rho_{l}^{-1} \Delta
P^{-3}\right)^{1/2}$, so that dimensional analysis together with
equation (\ref{ff}) readily yields \be
d_j/d_o=k_d(Q_l/Q_o)^{1/2}\label{d1} \ee where
$k_d=(8/\pi^2)^{1/4}$. This expression provides a first-order
approximation to the jet diameter (asymptotically true for $Re
\rightarrow \infty$) as far as $d_j>>d_o$ (negligible surface
tension). The ratio $Q_l/Q_o$ spans the whole domain from jetting
to dripping -where $d_j$ becomes comparable to $d_o$-. Jet
diameters and flow rates comparable to $d_o$ and $Q_o$,
respectively, lead to absolute instability, where the
characteristic velocity of upstream capillary waves $O(\sigma
\rho_l^{-1} d_j^{-1})^{1/2}$ -a product of surface tension-
becomes of the order of the downstream convective velocity,
$Q_l/d_j^2$. Besides, Eq. (\ref{d1}) is explicitly independent of
the orifice diameter $D$, an illustration of the jetting regime
independently of its forcing geometry.

Similarly, viscous effects can be weighted by defining a viscosity-related length
$d_\mu=(\mu_l^2\rho_l^{-1} \Delta P^{-1})^{1/2}$ and flow rate
$Q_\mu=(\mu_l^4\rho_l^{-3}\Delta P^{-1})^{1/2}$. Using these and equation (\ref{ff}), an
entirely analogous expression is obtained in the limit of dominant inertia, i. e. when
$d_j>> d_\mu$ and $Q_l>>Q_\mu$: \be d_j/d_\mu =k_d (Q_l/Q_\mu)^{1/2}.\label{d2}\ee This
equation expresses the jet diameter as compared to a limit where viscous effects become
important. Again, jet diameters and flow rates comparable to $d_\mu$ and $Q_\mu$,
respectively, amount to non-negligible viscous effects and significant departures from
predictions (\ref{d1}) or (\ref{d2}).

A third expression can be obtained in terms of the orifice diameter $D$, and the maximum
liquid flow rate that can be ejected through the orifice for a given $\Delta P$ in the
absence of viscous effects: $Q_{max}=(\pi^2/8)^{1/2}Q_m$, where -naturally- $Q_m=
(D^4\Delta P \rho_l^{-1})^{1/2}$ is obtained from dimensional analysis using $\{\Delta P,
D,\sigma\}$. Using (\ref{ff}) anew, one has: \be
d_j/D=(Q_l/Q_{max})^{1/2}=k_d(Q_l/Q_m)^{1/2}\label{d3}.\ee This alternative expression
reflects how close the experiment is from a situation where the entire orifice section is
filled with liquid: it provides information -from continuity arguments- on the fraction
of the orifice cross section occupied by the liquid jet.

Each of the above three expressions (\ref{d1}), (\ref{d2}) and (\ref{d3}) amount to
interesting but partial pictures of the particular working conditions of our system in a
given flow situation. Taken as a whole, they provide a more complete picture on the FF
jetting conditions. Some corrections can be obtained for several neglected
effects.\cite{advmat06}

\subsubsection{Correction for surface tension effects}

The liquid surface tension reduces the effective pressure drop
$\Delta P_l$ in the liquid stream as \be \Delta P_l=\Delta P -
2\sigma/d_j.\ee Consequently, the jet velocity decreases and its
diameter increases accordingly. The resulting expression for the
non-dimensional jet diameter $d_j/d_o$, neglecting third order
terms proportional to $O(d_o/d_j)<<1$, reads: \be
d_j/d_o=(8/\pi^2)^{1/4}(Q_l/Q_o)^{1/2}+1/2.\ee In other words, the
second order correction of the jet diameter $d_j$ to account for
surface tension effects is asymptotically equal to $d_o/2$.

\subsubsection{Correction for liquid viscosity effects (extensional stresses)}

Assuming that the extensional viscous forces in the liquid are
smaller than inertia, the balance of the different terms of the
momentum equation, including the second order terms of the
expansion, leads to the following order of magnitude for the
correction to the first order diameter (\ref{ff}): \be
d_e=O\left[d_\mu\left(\fr{Q_{max}}{Q_l}\right)^{1/2}\right]\ee

\subsubsection{Correction for tangential stresses owing to the gas stream}

In the same way, the diameter correction (decrease) owing to the
momentum injected by the much faster gas stream through the jet
surface is of the order of: \be d_g=O\left(\fr{\mu_g U_g D}{\Delta
P}\right)^{1/2} \ee where $\mu_g$ and $U_g$ are the gas viscosity
and velocity. The latter is of the order of $U_g \sim
O\left(\Delta P/\rho_g\right)^{1/2}$, where $\rho_g$ is the gas
density.

The relative weight of these three corrections provides information on the importance of
the surface tension and the viscosity of the liquid and gas phases. Interestingly, for
most common solvents, these relative weights are of order unity. This happens to be the
case when measuring the relative importance of the surface tension and the gas tangential
stress effects for water focused by any gas at standard conditions. Therefore, since both
corrections are opposite, the best agreement with experimentally measured jet diameters
and numerical simulations is obtained, interestingly enough, using the first order
expression (\ref{ff}), or its alternative forms (\ref{d1}-\ref{d3}).

\subsubsection{Correction owing to the nozzle flow pattern}

The jet diameter as measured at the nozzle may also differ from the simplest theoretical
prediction given by Eq. (\ref{ff}) because of local flow effects. A complex but symmetric
structure develops owing to the coexistence of (i) a core potential flow and (ii) the
detachment of a radially convergent boundary layer at the inner lip of the nozzle. In any
real situation where the gas viscosity is \it non-zero \rm and the continuum hypothesis
holds, this flow pattern is not aptly described by the pure potential flow through a
round orifice given by Morse and Feshback\cite{MF53} (page 1294) for a stationary
discharge. The potential flow solution is characterized by an axial velocity distribution
with a minimum value at the axis, $v(r=0)=2Q_g/(\pi D^2)$ (half the average velocity
through the orifice), and an infinite value at $r=D/2$, $Q_g$ being the theoretical gas
flow rate discharged. The actual flow geometry is characterized by the well known \it
vena contracta \rm effect, a consequence of the radial momentum carried by the collapsing
potential flow, which slips at the nozzle border owing to the boundary layer. The \it
vena contracta \rm flow exhibits an axial velocity distribution which echoes the
potential flow solution, showing a local minimum velocity at the axis, and a maximum
value at the streamlines coming just from the outside of the boundary layer detached at
the orifice (see Fig. \ref{f2-9}). The immediate consequence of this particular flow
structure is that the transversal pressure gradients are negligible only sufficiently far
downstream of the inner lip of the nozzle: in fact, they become negligible at the axial
downstream station where the \it vena contracta \rm effect ends, i.e. where the
streamlines become almost parallel. It occurs relatively close to the inner orifice
plane, at an approximate $D/2$ downstream distance. From this point downstream (before
shear instabilities of the gas stream with the external environment develop), the gas
pressure can be considered almost constant, equal to the outside stagnation value. It is
at this point where the liquid jet diameter obtained from the numerical simulation should
be compared to the simplest prediction (\ref{ff}).

\subsection{Scaling of the recirculation zone}

\begin{figure}
\begin{center}
\resizebox{0.9\textwidth}{!}{\includegraphics{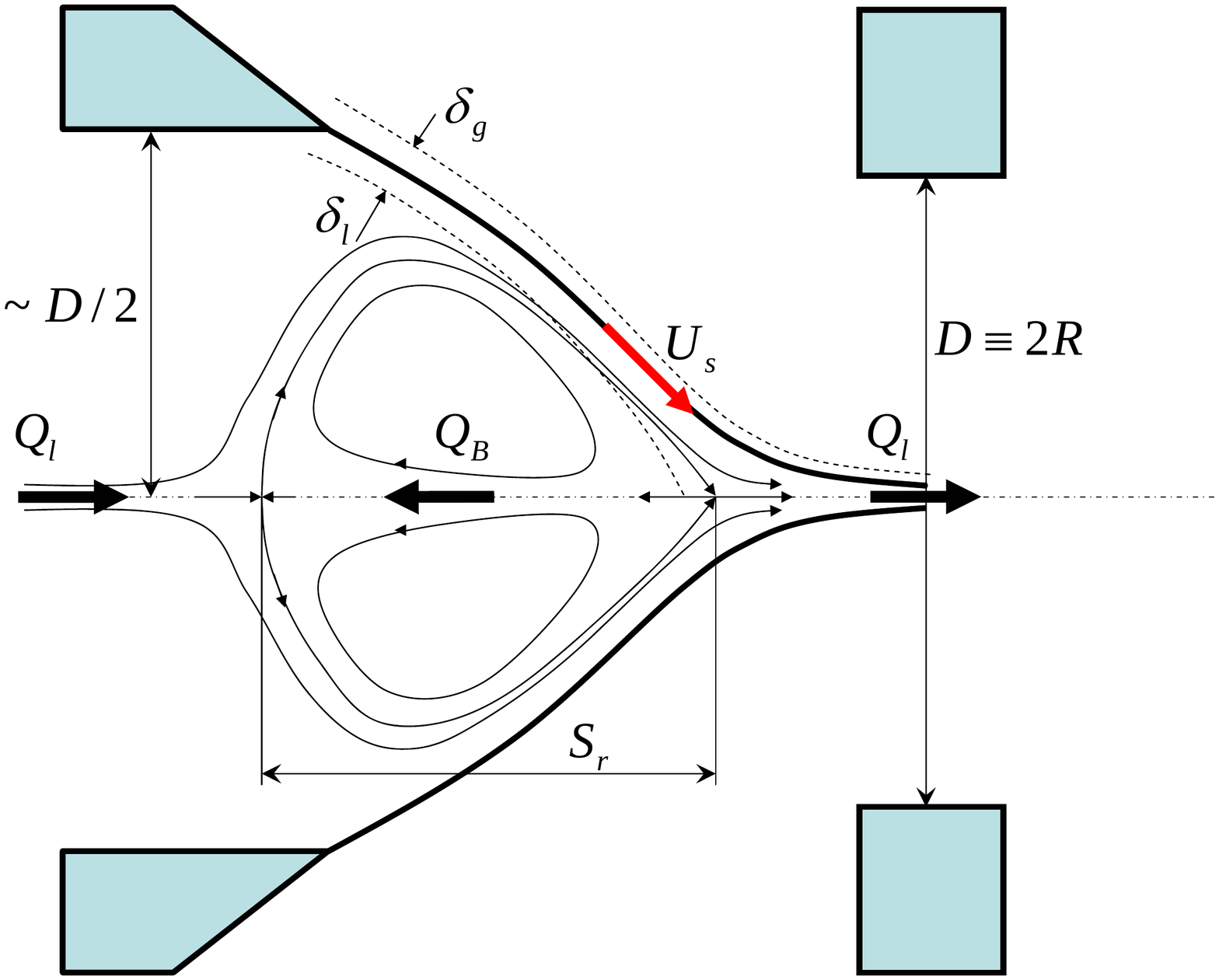}}
\end{center}
\caption{Sketch of the recirculation zone, showing boundary layers, cell size ($S_r$), and typical
velocities.} \label{f-r}
\end{figure}

For a given gas flow rate $Q_g$ and orifice diameter $D = 2 R$, the typical gas velocity close to
the meniscus surface can be estimated as $V=Q_g/(\pi R^2)$. Given the small $\rho=\rho_g/\rho_l$
values in liquid jets focused by gas, liquid velocities are much smaller than $V$ everywhere. As
the liquid approaches the neck, the boundary layer will collapse (Fig. \ref{f-r}). This implies
that at least a liquid flow rate \be Q_R \sim U_s \delta_l^2\label{QR}\ee would be drawn into the
jet in the absence of recirculation ($U_s$ is the velocity of the interface, that can be obtained
from $V$, and the densities and the viscosities ratios\cite{Gan07PRL}). On the contrary, whenever
$Q_l < Q_R$, part of $Q_R$ must have been recirculated back into the meniscus (Fig. \ref{f-r}).
Therefore $Q_R$ can be interpreted as the minimum flow rate for no recirculation (scaled as
$Q_r=Q_R/Q_g$).

The boundary layer in the liquid meniscus is confined. It grows along the cone during lengths
comparable to $R$ (the orifice radius) till the apex of the meniscus is reached. In this area, the
gas speed gradients are steep: \be \delta_l \sim (\mu_l R/\rho_l U_s)^{1/2}.\ee Whenever there is
recirculation, the peripheral boundary layers merge at the meniscus apex and give rise to a jet,
whose initial radius at the neck will accordingly be of the same order. In the absence of liquid
emission, maximum recirculation will be observed. Experimentally, however, a dripping instability
will occur before reaching this limit. In the opposite case (no recirculation), the boundary layers
do not merge, and an inviscid core should be observed at the neck. The threshold flow rate for
recirculation can therefore be estimated as $Q_R \sim U_s \delta_l^{2}$, a result which happens to
be independent of the gas velocity. In effect, by definition of the meniscus boundary layer, the
viscous stress $\mu_{l} U_s / \delta_l^{2}$ must be of the same order as the momentum convection
$\rho_{l} U_s^{2} / R$, so that, interestingly:
 \be Q_R \sim R \frac{\mu_l}{\rho_l}\Longrightarrow Q_r\sim \frac{\rho}{\mu Re}\label{Q-R}\ee
This scaling is fully confirmed by the numerical simulations: the
values of $Q_r\cdot Re$ are 0.6768 for case 1 and 0.6596 for case
2, deviating by less than 2.6\% from the scaling predictions.

\begin{figure}
\centerline{\includegraphics[width=0.80\textwidth]{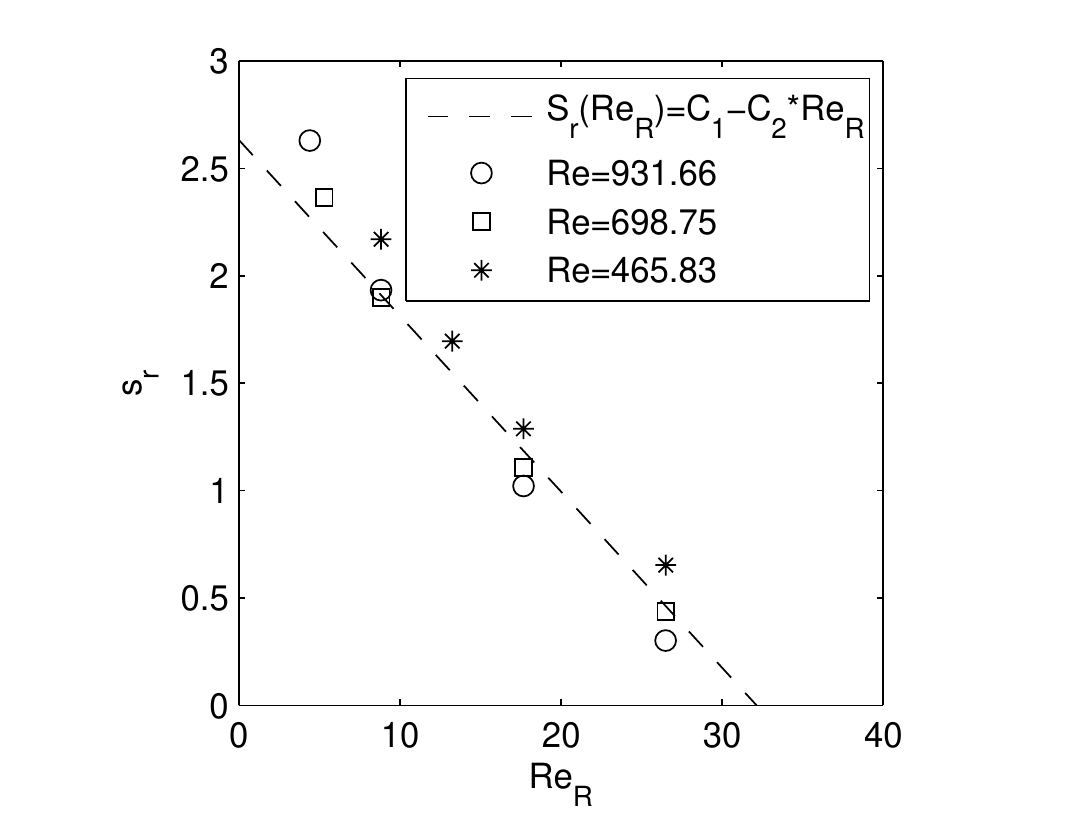}}\caption{Recirculation cell size $s_r$
as a function of $Re_R$: dots, squares and stars are obtained by numerical simulation; the line is
a theoretical prediction resulting from dimensional arguments. An additional series of simulations
have been performed for an intermediate gas flow condition ($Re=698.75$ and $We=18.31$) to assess
the validity of the scaling proposed: note the good degree of collapse obtained. The small
deviations can be attributable to the small differences in the geometry of the cone for different
gas flow conditions.}\label{final}
\end{figure}

Assume now the recirculation cell to be $S_r$ in axial length. The backflow $Q_B=Q_R-Q_l$ will come
to rest within a length of the order $S_r$. In this length, viscous momentum diffusion should slow
down the flow and deflect both the incoming flow injected by the feeding tube and the recirculated
flow at the axis (Fig. \ref{f-r}). Thus, viscous and inertia forces should balance within that
length $S_r$: in other words, the liquid Reynolds number associated to axial lengths of order $S_r$
should be of order unity so as to deflect the unidirectional flow issuing from the feed tube
(Hagen-Poiseuille). This is in analogy to the entry length or exit length in laminar pipe flow. Two
cases need to be considered, depending of the relative size of the cell compared to the feed tube
radius $R_1$:
\begin{itemize}
\item When $S_r<R_1$, viscous stress, of the order $O(\mu_l Q_B
S_r^{-3})$, balances inertia, $O(\rho_l Q_B^2 S_r^{-4})$, which
leads to $S_r\sim \rho_l Q_B \mu_l^{-1}$.
\item When $S_r>R_1$,
viscous stress, $O(\mu_l Q_B R_1^{-4})$, balances inertia,
$O(\rho_l Q_B^2 R_1^{-4}S_r^{-1})$, leading again to $S_r\sim
\rho_l Q_B \mu_l^{-1}$.
\end{itemize}
Interestingly enough, again, the length of the recirculation flow
is independent of the gas flow for any given geometry. The latter
scaling can be expressed in non-dimensional terms as: \be s_r
\equiv S_r/R\sim \mu_l(Q_R-Q_l)\rho_l^{-1}R^{-1}.\ee Using
equation (\ref{QR}), one may write: \be s_r=C_1-C_2
Re_R,\label{srn}\ee where $Re_R=\rho_l Q_l/(\mu_l R)$ is a
Reynolds number of the liquid flow, and $C_1$ and $C_2$ are
constants which depend on the geometry only (i.e., $R_1/R$, $H/R$,
etc.). In our case, we have represented all our measured $s_r$
values from numerical simulations versus $Re_R$ in Fig.
(\ref{final}). Linear fitting to all points leads to $C_1=2.636$
and $C_2=0.0819$ with a correlation coefficient of $94.4\%$.
Equation (\ref{srn}) can be expressed in terms of $Q_r-Q$ and $Re$
as well, as: \be s_r=k \mu\rho^{-1}(Q_r-Q),\ee where $k$ is again
a constant depending on the geometry only, in full agreement with
expression (\ref{RL}), as anticipated by experiments.

\section{Experiments}

\begin{figure}
\includegraphics[width=0.40\textwidth]{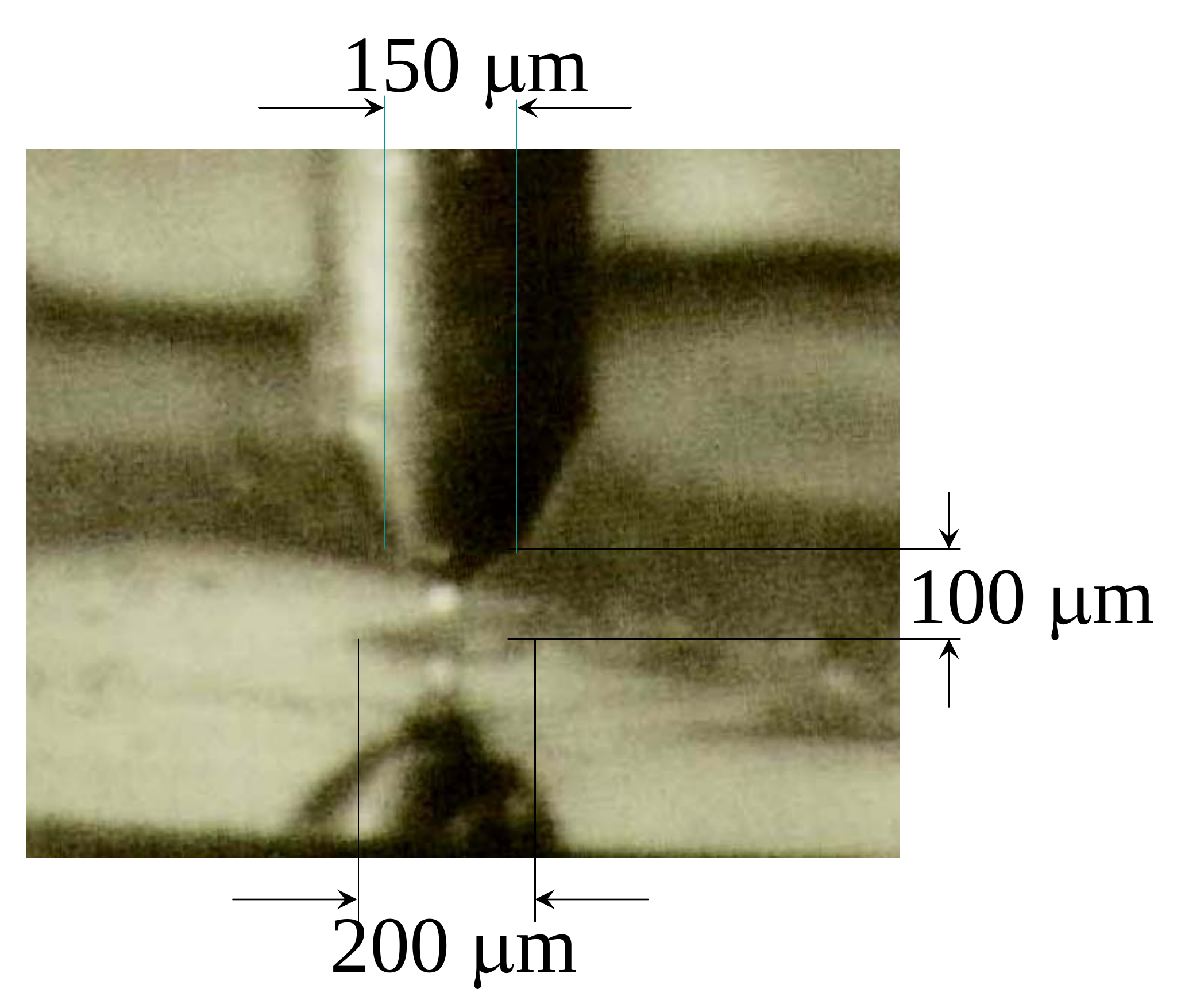}
\includegraphics[width=0.40\textwidth]{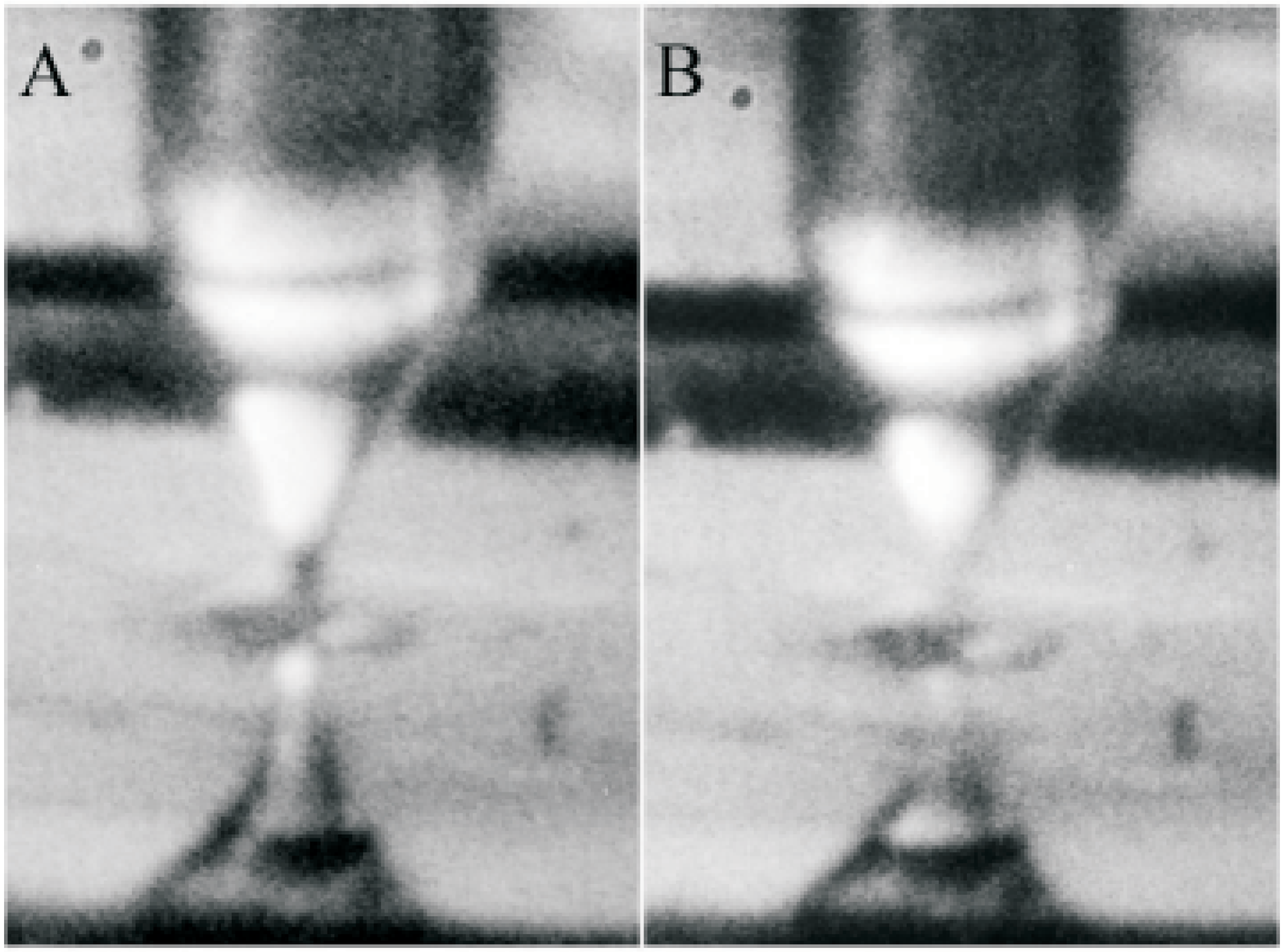}
\caption{(Left) Experimental tube-orifice set up as numerically
simulated in this work ($D=200 \mu$m, $D_1=150 \mu$m, $H=100
\mu$m; here, $\Delta P=10 KPa$, $Q_l=3$ mL/h). (Right) Photographs
of experimental conditions with twice the distance from the
feeding tube to the exit orifice, using a different tube material
(fused silica): (A) jetting ($D=200 \mu$m, $D_1=150 \mu$m, $H=200
\mu$m, $\Delta P=30 KPa$, $Q_l=6.1$ mL/h) and (B) dripping (as in
A, with $Q_l=2.8$ mL/h).} \label{exp1}
\end{figure}

In the following, we provide experiments corresponding to the same
local geometrical parameters as in cases 1 and 2 in the vicinity
of the exit orifice. The basic flow focusing chamber is a box
consisting of five aluminum faces and one clear methacrylate face.
It is 5 cm by 5 cm by 5.65 cm, with its longest side along the
capillary/orifice axis. The chamber is situated with the
methacrylate face horizontal and pointing upwards, the capillary
being located parallel to this face. The orifice is made in a
stainless steel orifice disk attached to the box side,
perpendicular and opposite to the capillary. The disk is 4.0 mm in
diameter with a thickness of 75 $\mu$m and an orifice of diameter
0.200 mm. Both the air tube and the capillary enter through the
face opposite the orifice. After the capillary tube has been
aligned with the orifice, the distance $H$ from the tube to the
orifice can be simply adjusted by carefully sliding the capillary
in its housing on the opposite face to the orifice disk. $H$ is
measured with a microscope through the methacrylate face. Fig.
\ref{exp1} shows some views of the feeding tube-orifice setup as
seen through the thick methacrylate window (inevitable liquid
spills leave behind some debris on the inner face of the window
causing a blurred image). In particular, Fig. \ref{exp1} (left)
shows the geometry numerically simulated in this work.

\begin{figure}
\centerline{\includegraphics[width=130mm]{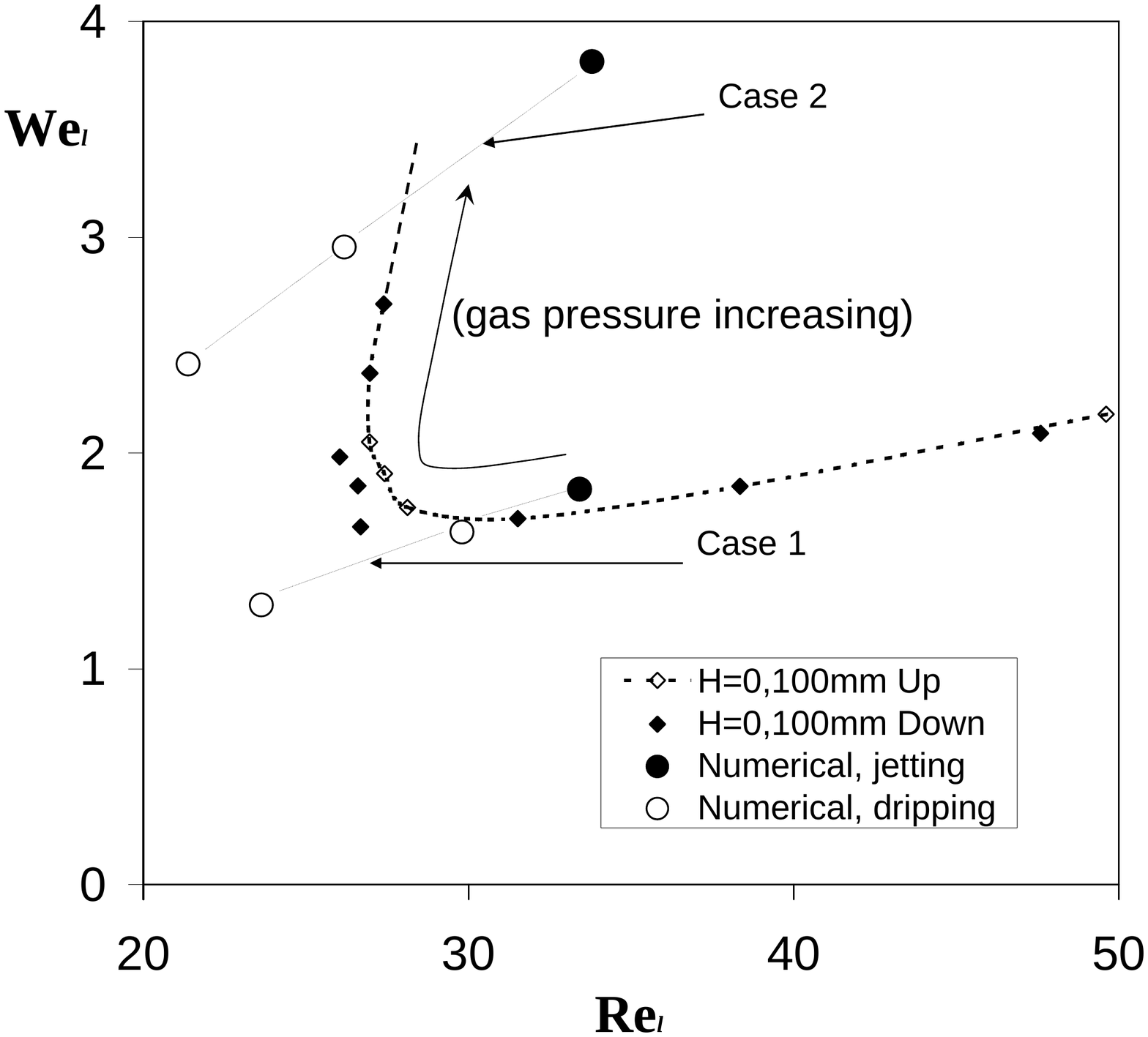}} \caption{Jetting -dripping transition in the
$\{Re_l, W\!e_l\}$ plane. Diamonds: experimentally determined conditions (filled symbols: liquid
flow rate decreasing -``Down''; open symbols: liquid flow rate increasing-``Up''. In most cases,
both ``Up'' and ``Down'' points coincide). Circles: numerically tested conditions. Filled circles:
jetting conditions. Open circles: dripping conditions.} \label{f2}
\end{figure}

After setting $H$ and ensuring that the capillary is perfectly coaxial with the nozzle orifice, the
pressure is set using a pressure gauge and a pressure meter. A water flow rate is then supplied
using a syringe pump (Cole-Palmer 74900 Series) with a 20 ml syringe. The system is given
sufficient time to relax until either a characteristic steady or unsteady flow is present. This can
be checked by illuminating the jet that exits the orifice or by looking at the meniscus when the
distance $H$ is 0.100 mm or greater. Unsteady jet flow appears very faint to the naked eye and
contains thin streaks of water along with large scattered spray. This is in significant contrast to
steady jet flow, which has bright illumination as a result of a finer, concentrated stream with
uniform characteristics. In experiments where the meniscus was visible, it was also possible to
discriminate steady versus unsteady flow (see Fig.\ref{exp1} (right) A -jetting- and B -dripping-),
in perfect correlation with the spray observations: a steady meniscus had sharp edges and a clear,
unwavering glass-like appearance (Fig. \ref{exp1} A, see steady jet reflected in the metal plate),
while an unsteady meniscus had blurred edges and flickered (Fig. \ref{exp1} B, no jet is visible at
all). Both the jet test and meniscus test displayed clear and abrupt transitions between the two
states. Once unsteady flow is established for a given pressure, the rate determined by the syringe
pump is increased in steps of 0.1 ml/hr. After each flow rate increase, a 30 seconds waiting period
was established, so as to ensure that the system had relaxed and all the readings were accurate.
This period has been chosen after it was found that $15 s$ was not enough to observe fluctuations
in the system: occasionally a steady regime would revert back to an unsteady one after the $15 s$
period. The $30 s$ delay has proven long enough to accurately characterize the flow. Accordingly,
the rate was increased until the unsteady jet sharply transitioned to a steady one; at this point,
the flow rate was read from the syringe pump and recorded as the minimal flow rate (increasing, or
``up''; Fig. \ref{f2}). Keeping this same steady flow rate the process is then reversed to find the
minimum decreasing (or ``down'') flow rate (steps of 0.1 ml/hr and intervals of 30 $s$ until an
unsteady regime developed). When the flow became unsteady, a rate 0.1 ml/hr above the reading on
the syringe pump was recorded, since the rate which produced the last steady flow (i.e. minimum
flow rate) was one step (0.1 ml/hr) higher. The resulting value was recorded as the minimal
(decreasing, or ``down'') flow rate. This process is repeated for varying pressures and distances
of $H$ to get an accurate mapping of minimal jetting flow rates as a function of varying geometry
and flow conditions. Following this procedure, we collected the experimental data plotted in Fig.
\ref{f2} for $H/R=1$. The gas (air) pressure $\Delta P$ increases as indicated by the arrow.

Six conditions numerically tested for cases 1 and 2 are plotted in Fig.\ref{f2}. In order
to make our results readily translatable in most of the capillary jet stability
literature (which uses the jet radius as a characteristic length), we may introduce
liquid Reynolds and Weber numbers consistent with previous definitions and using scaling
law (\ref{ff}): \be Re_l=\left(\fr{2}{\pi^2}\right)^{1/4}\left(\fr{\rho_l^3 Q_l^2\Delta
P}{\mu_l^4}\right)^{1/4} \,,\,\, We_l=\left(\fr{\pi^2}{8}\right)^{1/4}\left(\fr{\rho_l
Q_l^2\Delta P^3}{\sigma^4}\right)^{1/4} \ee As it follows from the plot, using these
definitions, jetting or dripping conditions are accurately predicted by the numerical
model. This lends additional support to the use of full VOF simulation analysis in flow
focusing systems.

\section{Conclusions}

The cone-jet geometry associated with flow focusing has been
handled by a diversity of tools, numerical, experimental and
theoretical. Order-of-magnitude estimations follow from
dimensional arguments: such procedures contribute a valuable
theoretical framing and provide the scaling criteria for data
representation. Analytical approaches are generally based on the
consideration of a perfectly cylindrical infinite jet, a
simplification that ignores the influence of the meniscus (a
source of instability) and the role of streamline convergence or
divergence in the jet. Experiments are burdened by the diversity
of influencing parameters and visualization difficulties
associated with the small scale of the meniscus and jet.

In this paper, experimental results are backed up by a numerical
simulation based on VOF elements. Numerical schemes allow a more
systematic exploration of the parametric influence. In addition,
the shortcomings of theoretical models (unavoidable in a situation
where the geometry of the fluid domain is complex, as in a
cone-jet flow pattern) are overcome, and a detailed description of
the streamlines can be readily obtained.

The key results of the above exploration are the following:

\begin{itemize}
\item The theoretical scaling leading to jet diameter estimates is
confirmed by the simulation. The expressions for flow focusing scales, notwithstanding their
simplicity, are therefore to be considered a reliable shortcut for the prediction of jet
dimensions. \item The complete sequence from meniscus growth to jet emission (jetting regime) and
to the sequential filling of drops (dripping regime) is portrayed in detail. \item The
jetting-dripping transition is documented in detail, both by experiment and simulation. A
two-branch structure is observed in the plot, showing the simultaneous influence of the jet and the
meniscus as instability sources. Incipient dripping (Fig. \ref{f2-14b}) is shown to give rise to
highly irregular fluctuations; while fully developed dripping (Fig. \ref{f2-14}) produces perfect
cycles of drop detachment. \item A recirculation cell is identified in the jetting regime at the
meniscus tip. This occurrence appears to be linked to intensive forcing by the gas sheath, leading
to high interface velocity along the meridians; the issuing jet is unable to convey all of the
mobilized flow, so that a return flow around the axis is observed. The recirculation cell grows as
the liquid flow rate is reduced: eventually, dripping conditions are reached. Similar recirculation
cells have been observed in electrospray cones, under thread emission, and in liquid-liquid
two-dimensional flow focusing, assisted or not by a surfactant \citep{SB2006,AnnaMayer06,BG98}. All
of the recirculation instances reported thus far appear to share a common attribute: strong
interfacial forcing, either electric, capillary or hydrodynamic. \item A reliable scaling is
provided, identifying the parametric conditions where recirculation is to be expected and
estimating the size and flow rates of the cell.
\end{itemize}

A key feature in the flow pattern explored is the recirculation
cell, and its conceptual link to the merging of the boundary
layers which grow from the meniscus edge and fuse together at the
neck of the jet. Controllable recirculation is an extremely
attractive feature, providing adjustable residence times within a
very simple flow setup. The cell can be viewed as a flow trap or
reactor, where biosynthesis or chemical operations take place in a
protected environment; the liquid flow rate can be increased to
flush the recirculation products.

An additional focus deals with the peculiarities of the jetting
and dripping regimes under the influence of a co-flowing sheath
current. The aspect of the jet and droplet train and the dynamics
of the meniscus (an indicator of dripping) are a contribution to a
problem whose complexity forbids a global theoretical approach.

\section{Acknowledgments}
This work has been supported by the Spanish Ministry of Science and Education, project
number DPI2004-07197, and partially supported by European Commission through grant
COOP-CT-2005-017725. Thorough discussions of one of the authors (AMGC) with Dr. Joan
Rosell are warmly acknowledged.

\newpage

\bibliographystyle{unsrtnat}

\begin{thebibliography}{56}
\bibitem[Flu()]{Fluent}
For convenience and with the aim of making our results readily reproducible
  from others, in this work we have used the well tested commercial solver
  FLUENT v 6.3. Mesh generation is aided by the software GAMBIT in FLUENT v
  6.2.
\end{thebibliography}

\end{document}